Proceedings of the
**6**<sup>**th**</sup> **International Workshop on**
**Multiple Partonic Interactions at the LHC**

# MPI@LHC 2014

3-7 November, 2014
Kraków, Poland

# Organizing Committee

### International Advisory Board

Paolo Bartalini - China Central Normal University  
Jonathan Butterworth - UC London  
Markus Diehl - DESY  
Livio Fanò - University of Perugia, INFN  
Rickard Field - University of Florida  
Hannes Jung - DESY, University of Antwerp  
Judith Katzy - DESY  
Frank Krauss - IPPP Durham  
Krzysztof Kutak - IFJ PAN Krakow  
Stefano Lami - CERN  
Michelangelo Mangano - CERN  
Arthur Moraes - Centro Brasileiro de Pesquisas Físicas  
Andreas Morsch - CERN  
Zoltan Nagy - DESY  
Giulia Pancheri - INFN Frascati Nat. Lab.  
Michael Schmelling - MPIK Heidelberg  
Torbjorn Sjöstrand - Lund University  
Peter Skands - CERN, Monash University  
Mark Strikman - Pennsylvania State University  
Antoni Szczurek - IFJ PAN Krakow  
Daniele Treleani - INFN Trieste  
Pierre van Mechelen - University of Antwerp  
Nick van Remortel - University of Antwerp  

### Local Organising Committee

Krzysztof Kutak (chair)  
Antoni Szczurek (chair)  
Rafał Maciuła  
Rafał Staszewski  
Maciej Trzebiński  

# List of Authors

R. Astalos (Comenius University, Bratislava), S. Baranov (Russian Academy of Sciences), P. Bartalini (Central China Normal University CCNU), I. Belyaev (Institute for Theoretical And Experimental Physics, Moscow), Ch. Bierlich (Lund University), B. Blok (Technion, Israel), A. Buckley (University of Glasgow), F. A. Ceccopieri (Universite de Liege), I. Cherednikov (University of Antwerp), J. R. Christiansen (Lund University), D. Ciangottini (CERN/INFN), M. Deak (IFIC/CSIC, Universitat de Valencia), B. Ducloue (University of Jyvaskyla), R. Field (University of Florida), J. R. Gaunt (DESY, Hamburg), K. Golec-Biernat (IFJ PAN, Krakow/Rzeszow University), L. Goerlich (IFJ PAN, Krakow), A. Grebenyuk (Université Libre De Bruxelles), O. Gueta (Tel Aviv University), P. Gunnellini (DESY, Hamburg/Universiteit Antwerpen), I. Helenius (Lund University), H. Jung (DESY, Hamburg), D. Kar (University of Glasgow), O. Kepka (Acad. of Sciences of the Czech Rep.), M. Kłusek-Gawenda (IFJ PAN, Krakow), A. Knutsson (University of Antwerp), P. Kotko (Penn State University), M. W. Krasny (Univ. Paris VI), K. Kutak (IFJ PAN, Krakow), E. Lewandowska (IFJ PAN, Krakow), G. Lykasov (Joint Institute for Nuclear Research, Dubna), R. Maciuła (IFJ PAN, Krakow), A. M. Moraes (Brazilian Center for Physics Research), T. Martin (University of Warwick), G. Mitsuka (Universita e INFN), L. Motyka (Jagiellonian University), M. Myska (CTU, Prague), J. Otwinowski (IFJ PAN, Krakow), T. Pierog (KIT/IKP), V. Pleskot (Charles University in Prague), M. Rinaldi (INFN, Perugia), W. Schäfer (IFJ PAN, Krakow), A. Siodmok (CERN/IFJ PAN, Krakow), T. Sjöstrand (Lund Univeristy), A. Snigirev (SINP MSU), A. Staśto (Penn State University/IFJ PAN, Krakow), R. Staszewski (IFJ PAN, Krakow), T. Stebel (Jagiellonian University), M. Strikman (Penn State University), A. Szczurek (IFJ PAN, Krakow/Rzeszów University), D. Treleani (University of Trieste), M. Trzebiński (IFJ PAN, Krakow), H. van Haevermaet (DESY, Hamburg/University of Antwerp), A. van Hameren (IFJ PAN, Krakow), P. van Mechelen (University of Antwerp), W. Waalewijn (University of Amsterdam/Nikhef), W. Y. Wang (National University of Singapore),

# Contents

# Preface

The Sixth International Workshop on Multiple Partonic Interactions at the Large Hadron Collider (MPI@LHC2014) took place at the Polish Academy of Arts and Sciences (PAU) in Kraków, from 3 to 7 November 2014.

MPI are often crucial for interpreting results obtained at the Large Hadron Collider (LHC). The quest for a sound understanding of the dynamics behind MPI - particularly at this time when the LHC is due to start its "Run II" operations - has focused the aim of this workshop. MPI@LHC2014 concentrated mainly on the phenomenology of LHC measurements whilst keeping in perspective those results obtained at previous hadron colliders. The workshop has also debated some of the *state-of-the-art* theoretical considerations and the modeling of MPI in Monte Carlo event generators.

The topics debated in the workshop included:

- Phenomenology of MPI processes and multiparton distributions;

- Considerations for the description of MPI in Quantum Chromodynamics (QCD);

- Measuring multiple partonic interactions;

- Experimental results on inelastic hadronic collisions: underlying event, minimum bias, forward energy flow;

- Monte Carlo generator development and tuning;

- Connections with low-x phenomena, diffraction, heavy-ion physics and cosmic rays.

In a total of 57 plenary talks the workshop covered a wide range of experimental results, Monte Carlo development and tuning, phenomenology and dedicated measurements of MPI which were produced with data from the LHC's Run I. Recent progress of theoretical understanding of MPI in pp, pA and AA collisions as well as the role of MPI in diffraction and small-x physics were also covered. Two discussion sessions focused on *Double Parton Scattering and the Underlying Event* and on *Small-x and Collisions with Nuclei* were also held at the workshop and debated how the lessons learned from the LHC Run I analyses could help the community to plan ahead for the incoming Run II.

The workshop forstered close contact between the experimental and theoretical communities. It provided a setting to discuss many of the different aspects of MPI, eventually identifying them as a unifying concept between apparently different lines of research and evaluating their impact on the LHC physics programme.

We want to thank the International Advisory Board for the MPI@LHC Workshop Series and the Working Group Conveners who contributed so effectively to the preparation of an interesting and successful scientific program for the workshop. We must also express our deepest gratitude to our sponsors: Institute of Nuclear Physics - Polish Academy of Sciences; National Science Centre - Poland; Krakow to business - KRK>2B. It is also a pleasure to thank the Local Organising Committee, Krzysztof Kutak (chair), Antoni Szczurek (chair), Rafal Maciula, Rafal Staszewski, Maciej Trzebinski and Jolanta Mosurek for their tireless work before and during the workshop.

In the following, the reader will find summary reports of plenary talks presented at the workshop. These contributions are organized in "chapters" corresponding to the working groups which were part of the scientific program.

All the presentations are available at: *https://indico.cern.ch/event/305160/overview* .

*Arthur M. Moraes*

Chapter:
# Underlying Event

# Outlook on Minimum Bias and Underlying Events

H. Jung

When LHC started in 2009/2010 none of the predictions for the charged multiplicity $dn/d\eta$ nor for the underlying event as a function of the transverse momentum of a leading track or jet was agreeing with the measurements. Since then, huge progress has been made in the modeling and understanding of minimum bias and the underlying event structure. Now, just before LHC restarts at $\sqrt{s} = 13$ TeV, all relevant models are able to describe the main features of the events at $\sqrt{s} = 7(8)$ TeV, after a big effort of tuning the free parameters.

However, different tunes exist for different observables: tunes for minimum bias events, for underlying events with jets and different parameters might be needed when running with higher order calculations. A big step forward was observed with the measurement of $dn/d\eta$ from CMS-Totem, which had a clean definition of the phase space in terms of particles in the forward detectors from Totem. A reasonable description of this measurement was achieved with the CMS underlying event tunes (without the need to further re-tune).

The latest up-to-date tunes from both ATLAS and CMS are able to describe reasonably well the existing measurements and the predictions to higher energies (for LHC run2 at 13 TeV) do not differ very much. However, ons should keep in mind, that a difference of 0.5 particles/$\eta$ in minimum bias events will lead to a differences of $\sim 20$ particles/$\eta$ at the high pileup conditions of run2. Therefore, the very first measurements of $dn/d\eta$ at 13 TeV will be very important.

Apart from a good description of the minimum bias and underlying events, a few conceptional issues have to be kept in mind:

- The detailed mechanism of the taming/saturation of the (mini-) jet cross section from large to very small $p_T$ and the connection to saturation physics is still not well understood.

- With the present tunes, it seems to be non-trivial to describe at the same time minimum bias, underlying events and hard double parton scattering. If the basic principles of multiparton interaction are appropriate at these different scales, a unified description over the full kinematic range is needed.

- With the present knowledge it is not clear, whether the same parameter tunes can be applied in NLO (or multi-leg) matched parton shower and hadronization calculations like POWHEG, MC@NLO or ALPGEN, MADGRAPH. For a unified description, this would be necessary.

- Experimental observables must be defined in a way, which is free from prejudice, i.e. without assuming the existence of multiparton interactions. The concept of multiparton interaction is directly interwoven with the concept of factorisation. If cross sections can be described with a different way of factorisation, then the role of multi-parton interactions might be very different.

# Outlook on MPI and MC models

A. Grebenyuk, D. Kar and A. Siodmok

**Introduction**

Multiple Parton Interaction (MPI) is among the least well understood phenomenon, but a proper modeling of it is necessary for measurements and searches. Monte Carlo (MC) event generators include different phenomenological models for MPI, and a significant progress in development of MPI models happened in response to LHC Run I data.

**LHC Run 1 Summary**

ATLAS, ALICE and CMS performed a host of measurements sensitive to different aspects of soft QCD, particularly MPI. They include the *minimum bias* (MB) charged particle distributions [1, 2, 3], transverse energy flow [4], and traditional underlying event (UE) measurements in different topologies [2, 6, 7, 8, 9, 3, 11], There have been other new measurements as well, but not all of them are corrected back to particle level, thereby reducing their usefulness.

The first charged multiplicity and leading-track underlying events measurements at 7 TeV, and surprisingly at 900 GeV showed incompatibility with all the pre-LHC Monte Carlo models, with the MPI part mostly *tuned* to 1.96 TeV Tevatron data. This led to a rapid retuning effort, described in the next section.

The tuning obviously is one part, where the soft-QCD resulted are utilised. On the other hand, many of the results have been interesting on their own accord. CMS used jet areas to estimate the contribution of underlying event [12], and ATLAS measured the standard deviation of the usual charged particle multiplicity or charged particle sum $p_T$ profiles (as a function of leading track $p_T$). A steady non-zero plateau-like shape demonstrated that UE effect cannot just be subtracted as an average value for all events. The ATLAS measurement with charged particle jets, varying the radius of charged particle jets showed the activity initially increases with increasing charged particle jet radius - pointing possibly to selection bias. Finally both CMS and ATLAS comparison of UE in dijet and $Z$-boson events showed very similar activity, confirming the similar behavior seen at CDF [13]. It was also realised that at very busy LHC environments, the traditional *transverse and trans-min regions* (with respect to the hard scattering direction in $\eta$ -$\phi$ space) is not just sensitive to UE, but receive contribution from additional radiation [9]. To mitigate the contamination, ATLAS looked at exclusive dijet topology, vetoing addition hard jets, which can be a promising way to gain sensitivity to MPI effects [8].

**Model development and Tuning**

The model development can be briefly summarised as:

- Pythia 6/8: had already well developed MPI model before Run 1 data. However, the disagreement with LHC data prompted the experiments to start a quick retuning effort and a tension between all three center-of-mass energies were observed. ATLAS and CMS, as well as Peter Skands came up with new Pythia 6 tunes which would describe the 7 TeV data better [14], termed the Perugia tune. While the Perugia series of tunes and tunes from CMS were aimed to describe both minimum-bias (MB) and underlying event (UE) data simultaneously, it became apparent that separate tunes for minimum bias and underlying event can result in individually better descriptions for either categories. While this option was certainly not the aesthetically pleasing one, ATLAS adapted this philosophy, and started producing separate MB and UE tunes. As the usage of Pythia 8 gained more popularity, ATLAS focused more attention to Pythia 8 tunes [15, 16, 17], and separately there were tunes by authors, Peter Skands (the Monash series [18]), and finally by CMS []. There has also been significant progress in the model development: $x$-dependent overlap function [20], MPI re-scattering [21], a lot of tunes [14, 18, 22] and most recently a new string formation beyond leading colour [23].

- Herwig++: the program includes model with hard [1] and soft [25] MPI scatters. The LHC MB data triggered development of colour reconnection (CR) models [3] and the authors provided new tunes [3, 4]. The most recent UE-EE-5 tune [4] is able to describe the double-parton scattering cross section [5, 6] and charged tracks underlying event data collected over the collision energy range from $\sqrt{s} = 300$ GeV to $\sqrt{s} = 7$ TeV.

- Sherpa: constructed and implemented a new MPI model SHRiMPS [30] based on KMR model, which is multi-channel eikonal model with partonic interpretation of pomeron. It has been released recently and should be tuned soon (to construct a new model it takes years!). Currently the model can be only used to described MB physics and the extention of the model to UE is ongoing.

Other programs which main aim is to describe cosmic rays or/and heavy ion collisions also made remarkable progress:

- EPOS which is used for both heavy ion interactions and cosmic ray air shower simulations released its LHC version [4] which provides successful description of LHC MB data.

- DIPSY [32] a MC generator primarily developed for simulation of MB in pp collisions and heavy ion physics, recently implemented so called rope model [33]. This model includes the effect of overlapping strings in the Lund Hadronisation Model.

With the advent of LHC, tuning of MC generators also saw an "industrial revolution". In addition to tuning by eye, started by Rick Field, use of automated tuning tool Professor [5] using the analysis toolkit Rivet [4] became popular. In parallel, platforms like mcplots [37] have made tune comparisons easy by using volunteer computer resources.

Historically tuning efforts have been by tuning the hadronisation model by LEP results, and MPI model by Tevatron and LHC results. Monash for example-tuned the hadronisation parameters by using LHC data as well, while the latest ATLAS effort [19] tuned the shower and MPI parameters simultaneously, coming up with a tune suitable for used in high $p_\text{T}$ processes, and matched setups.

**Open Issues and Run 2 Ideas**

Despite of all the progress and successes of the models there is definitively still scope for significant improvements in the future. There are aspects in data that our models cannot describe. For example, none of the models/tunes do a good job at describing the transverse energy flow in high rapidity range, as seen from both ATLAS and CMS results. There seems to be a conflict in describing the transverse energy flow in central rapidity regions in MB events with charged particle $p_\text{T}$ spectra. The particle content in events, more specifically the fraction of strange mesons and baryons is not modeled accurately by any generators.

Another example is that the observation that $\langle p_T \rangle$ increases with $n_{ch}$ suggests that there is some mechanism of correlating partons from different MPI systems. Models which have had some success in describing the $<p_T>(n_{ch})$ distributions have some method of stronger correlating MPI systems using colour reconnection (CR) (Herwig++ and Pythia) or hydrodynamic collective flow (EPOS). This part of modelling is definitively the biggest known unknown. It would be important task for future experimental and phenomenological studies to find ways to check which model (or maybe a superposition of both) is realized by the nature. It is worth to stress that the CR uncertainties are important to many precision measurements, they are especially important in the top mass measurements, see for details [34]. Unfortunately, CR uncertainties are often underestimated by the experimental collaborations (see T. Sjöstrand's talk). One way to address this problem is by dedicated experimental studies of CR in top events. Other examples where MPI uncertainty might be underestimated is DPS background to Higgs searches (as presented by W. Krasny) this problem should be also addressed by our community. The estimation of DPI/MPI cross section is tied to constructing generator specific MPI-on and MPI-off templates [2], where we have no evidence that the MPI modeling in any generator is accurate. CMS added the effective DPI cross-section as a tuning parameter, which results in a slightly improved description.

The first major challenge in Run 2 will obviously be to test the energy extrapolation of the models using the early (low pileup) 13 TeV data. In Fig. 1, it can be seen that some commonly used tunes have very different MPI $p_T$ cutoff at 13 TeV. The minimum bias measurements, and height of the underlying event plateau will be useful indicators of the models.

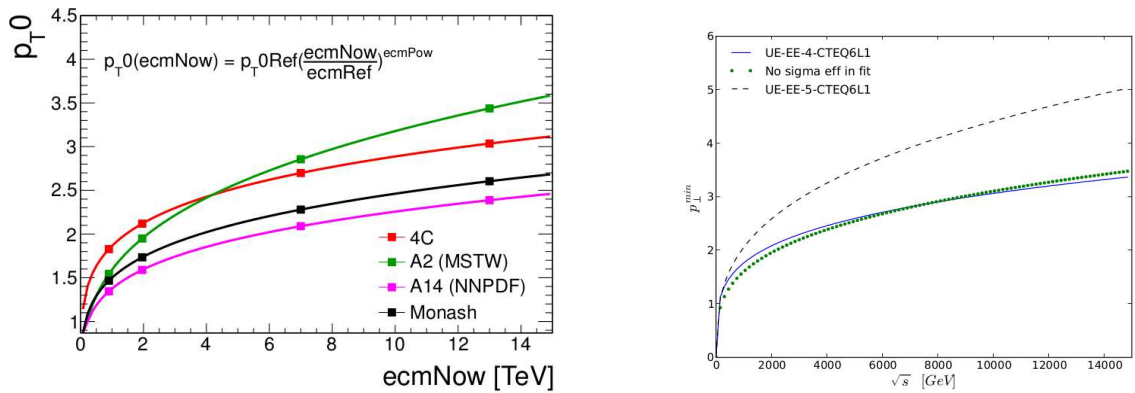

Figure 1: Energy Extrapolation of some Pythia8 (left) and Herwig++ (right) tunes

With more data, the focus will shift to modeling the MPI in high $p_T$ events. It will be essential to find observables or topologies which are less affected by extra jets, thereby enabling us to estimate MPI contribution to processes of interest in searches and measurements.

This also means that rather than leading order parton shower generators, generators with NLO or multileg matrix element interfaced with a parton shower generator can lead to a better description of data even for observables sensitive to soft effects.

There is also a lack of a uniform way to treat the model uncertainties. Experimentally this is often one of the dominant source of systematic uncertainty. Comparison between different generators or tunes has been the conventional way, but more recently ATLAS has advocated the eigentunes approach using the Professor system. There is still some way to go before we have a consistent recipe.

Run 2 data will help us to answer some of the questions, however some points discussed in Perugia 2008 meeting are equally relevant today

- How should the relative importance of various data be judged?
- When are discrepancies due to poor physics or to poorly documented/measured data?
- How can we avoid over-tuning, i.e. avoid forcing the model to fit the data even if the data contain physics not included in the model (which is often the demand of experimental collaborations)?

We end by remembering the warning from Perugia proceedings:
"In summary, the pride of recent successes should not blind us to the challenges ahead."

# MPI in PYTHIA

Torbjörn Sjöstrand

Department of Astronomy and Theoretical Physics
Lund University

The development of the MPI framework for hadronic collisions in PYTHIA was begun almost exactly 30 years ago. Most of the basic ideas of the early framework [1] have survived, and several have come to be generally accepted:
− multiple (semi)perturbative (mini)jet production as the origin for a unified description of hard jets, underlying events and minimum-bias events;
− screening as a mechanism to tame the divergence of the cross section in the $p_\perp \to 0$ limit, with a characteristic turnoff scale around 2 GeV;
− a $p_\perp$-ordered generation ensures that the PDFs of the hardest interaction can be the conventional ones, while the ones used for softer interactions need to be successively modified;
− an impact-parameter dependence gives a natural mechanism to obtain a pedestal effect; and
− colour reconnection is needed to explain the rise of $\langle p_\perp \rangle (n_{\mathrm{ch}})$.

Over the years this basic picture has been refined in numerous ways, including interleaved evolution of MPI, ISR and FSR, energy-dependent screening, more sophisticated multiparton PDFs, junction topologies, partonic rescattering, an $x$-dependent proton size, MPI in diffractive events, and more. These models have continuously been made publicly available, most recently in PYTHIA 8.2 [2]. Some have led to improvements in the description of data, while others have yet to make a visible impact. Much but not all data can be explained, leaving room for further improvements, and for the inclusion of additional effects, such as hadronic rescattering.

Potentially the biggest "known unknown" in the MPI framework is the role of colour reconnection (CR). The description of MPI, ISR and FSR leads to multiple colour confinement fields – strings – being spanned criss-cross in the event, often overlapping with each other. While this picture typically is formulated in the limit of an infinite number of colours, there is no indication that a more correct, but static, colour description would resolve the issue. Rather some kind of dynamical effects seem required, where the overlapping strings may reconnect, usually so as to reduce the total string length. This still leaves a big room for potential models.

The understanding of colour reconnection can have an impact on issues such as top mass reconstruction, where the experimental precision now is below 1 GeV. Since the PYTHIA 8.1 code only contained one CR model (unlike PYTHIA 6.4, which contained more), recently several new models were introduced [3]. These fall into two main classes. In one the top decay products only get the chance to reconnect after the rest of the event has been handled as usual, while in the other the top decay products are handled in the same way as other partons. The former class stealthily gives more freedom from experimental constraints, and therefore is convenient to explore worst-case scenarios. Experience with these models suggest that the CR top mass uncertainty could be as high as half a GeV, but with a rather asymmetric distribution.

Other models for CR have been presented by several other speakers at this meeting, whereof the one described by Jesper Roy Christiansen is also available in PYTHIA 8.2.

# A new Colour Reconnection model in PYTHIA

Jesper Roy Christiansen[1,2]

Department of Astronomy and Theoretical Physics
Lund University

The recent renewal of interest in colour reconnection (CR) at LHC is due to two observations: the model prediction for baryon production does not match the data (Fig. 2), and the CR uncertainty on the top mass has become one of the larger uncertainties. The last point was discussed in more detail in Torbjörn Sjöstrand's talk.

The new CR model (implemented in PYTHIA) builds on two main physics ideas:

- It uses the colour rules from SU(3) to determine which reconnections are allowed. Especially, the epsilon structure of the SU(3) group is also included, allowing the creation of junction structures (and thereby additional baryons).

- The $\lambda$-measure is used to determine which of the allowed colour configurations is chosen. Only reconnection which lowers the total string length (the $\lambda$-measure) is considered and thereby a local minimization is achieved.

A new beam remnant model following similar principles was also introduced in PYTHIA.

By tuning the amount of junction structures, the baryon production can be described simultaneously at both LEP and LHC (Fig. 2). Other models (also presented at this conference) describe the baryon production as well, however. All the models rely on the higher colour/energy density of LHC to provide the additional baryons needed, but the method to obtain them varies. A natural continuation is therefore to consider additional observables especially sensitive to the higher densities. One suggestion is to study the baryon production as a function of the multiplicity. With the correlation between multiplicity and energy/colour density, this observable can both probe the regime dominated by the new models as well as their onsets.

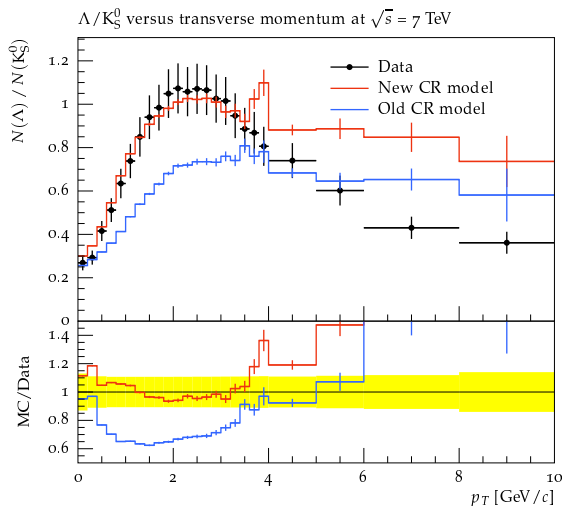

Figure 2: The $\Lambda/K_s^0$ $p_\perp$-distribution as recorded by the CMS collaboration [1]

Other suggested observables include the $\langle p_\perp \rangle$ vs mass, where the standard PYTHIA can not describe the slope even at lower centre-of-mass energies [2]. A corresponding measurement at LHC energies could shed light on its relation to energy/colour densities.

---

[1]Work supported by the MCnetITN FP7 Marie Curie Initial Training Network, contract PITN-GA-2012-315877.
[2]In collaboration with Peter Skands.

# Underlying events in EPOS 3.1xx

Tanguy Pierog

Karlsruhe Institute of Technology (KIT), Campus North, Karlsruhe, Germany

## Open problems

The measurement of underlying event activity has been initiated by the CDF collaboration more than ten years ago [1] to understand multiple parton interaction and test the capability of the event generators to reproduce real events. This analysis has been generalized to more recent collider data including the main LHC experiment ATLAS [2] and CMS [3]. With the minimum bias hadronic interaction model EPOS LHC [4] (dashed line), data are well reproduce below 5 $GeV/c$ (Fig. 3 right) with an active core producing collective flow but the description was not so good above 10 $GeV/c$. Without core (Fig. 3 left), simulations from EPOS LHC do not reproduce data at all because not enough hard scattering are present and flow is not there to increase the average transverse momentum. Using a preliminary version of EPOS 3 (full line) which include a new variable saturation scale [5] which provide more hard scattering, simulations without flow from collective hadronization (Fig. 3 left) reproduce very well the data. While with flow (Fig. 3 right) the description becomes very bad at high $p_t$ due to a too large energy loss in the high density medium which is probably overestimated here (based on PbPb collisions).

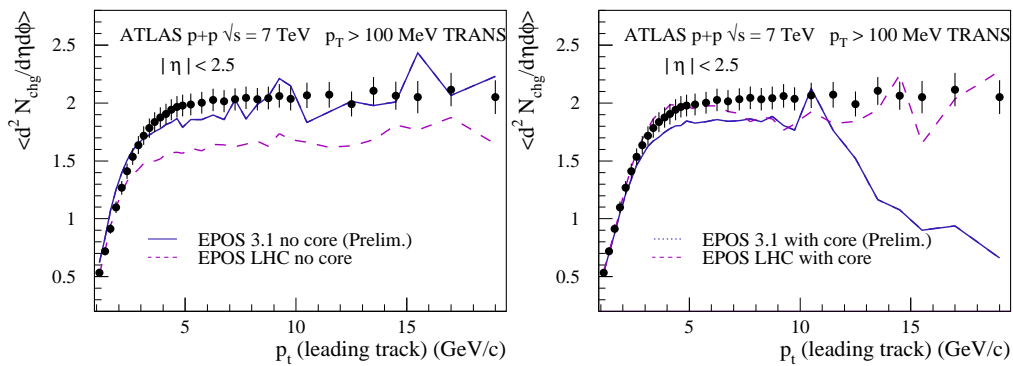

Figure 3: ATLAS measurement of charged particles with $p_t \geq 100$MeV produced in the transverse region [2] compared to EPOS LHC (dashed line) and EPOS 3.1 (solid line) simulations without (left) or with (right) core formation.

## Outlook

Collective flow (or color reconnection) are very important for the description of underlying events for leading track $p_t$ below 5 $Gev/c$. Collective flow treatment in EPOS 3 will be improve to describe both low and high $p_t$ particles.

# MPI in Herwig++

Andrzej Siodmok

CERN, PH-TH, CH-1211 Geneva 23, Switzerland

Models with hard [1] (similar to JIMMY [2] package) and soft component [3] of multiple partonic interactions as well as improved colour reconnection models [4] are available in Herwig++. The main parameters of the model are $p_{t,min}$ which sets a transition scale between hard and soft (non-perturbative) component, $\mu$ parameter which can be interpreted as the inverse radius of the proton and $p_{reco}$ the probablity of colour reconnection. The parameters are obtained from fits to previous MPI, UE and sigma effective data. The most recent UE-EE-5 tune [4] is able to describe the double-parton scattering cross section [5, 6] and charged tracks underlying event data collected over the collision energy range from $\sqrt{s} = 300$ GeV to $\sqrt{s} = 7$ TeV, see Fig. 4 and Fig. 5. The predictions for 14 TeV are shown in Fig. 5 (top sets).

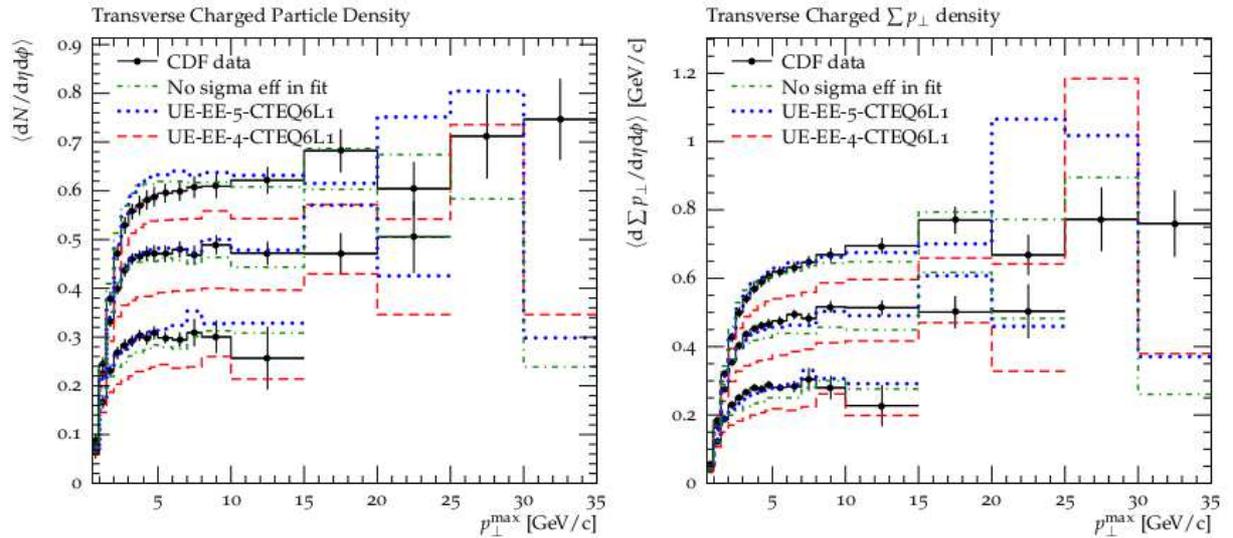

Figure 4: CDF data and model fits at at 300 GeV (lowest sets), 900 GeV (middle sets) and 1960 GeV (top sets), showing the mean multiplicity density (left panel) and mean scalar $p_\perp$ sum (right panel) of the stable charged particles in the "transverse" area as a function of $p_\perp^{\text{lead}}$.

There was also progress in improving the forward rapidity gap description in Herwig++ presented in a separate contribution by M. Myska.

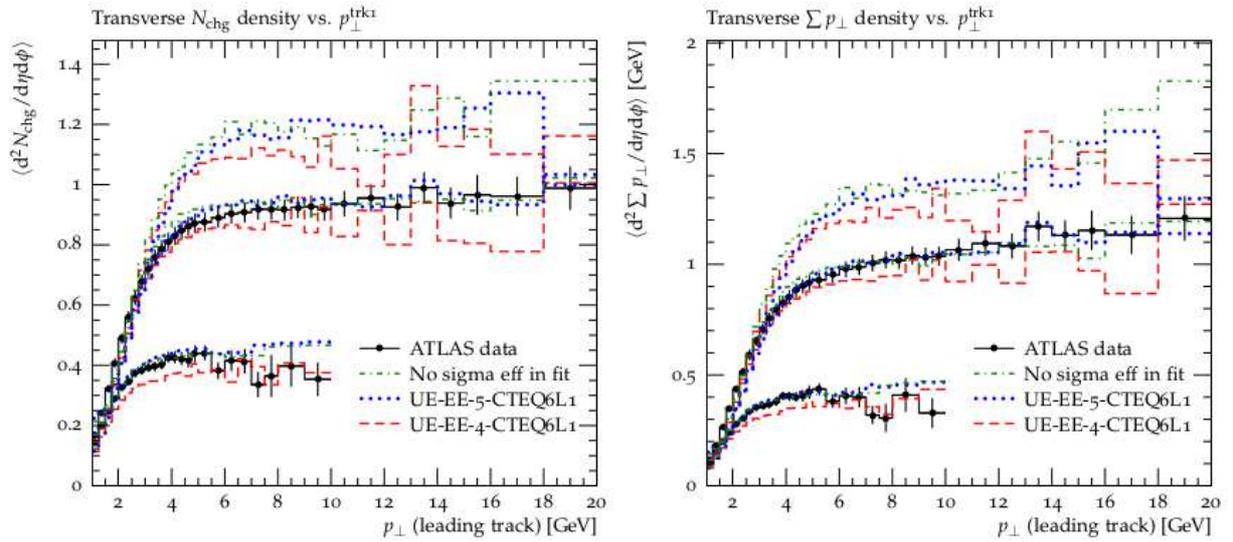

Figure 5: ATLAS data and model fits at 900 GeV (lowest sets) and 7 TeV (middle sets), showing the mean multiplicity density (left panel) and mean scalar $p_\perp$ sum (right panel) of the stable charged particles in the "transverse" area as a function of $p_\perp^{\text{lead}}$. The predictions for 14 TeV (top sets) are shown for comparison.

# Improving the forward rapidity gap description in Herwig++

Miroslav Myska[*,1], Jonathan Da Silva[2], Michael H. Seymour[2], Andrzej Siódmok[3]

[*] Speaker
[1] FNSPE, Czech Technical University in Prague
[2] School of Physics and Astronomy, University of Manchester
[3] Physics Department, CERN

Herwig++ [1] is a general-purpose Monte Carlo event generator for lepton and hadron collisions and it is successful in many directions, including standard underlying event measurements at the Tevatron and the LHC. However, Herwig++ did not turn out to correctly describe the forward rapidity gap ($\Delta\eta^F$) measurement [2]. $\Delta\eta^F$ is defined as the largest pseudorapidity difference between the most forward particle detected and the edge of the acceptance of the detector. Our investigation deals only with the non-diffractive component of the $\Delta\eta^F$ distribution, which should reach its maximum at $\Delta\eta^F \to 0$ and should fall exponentially down. However, Herwig++ produces large amount of events with very large gaps.

Since the minimal threshold on the transverse momentum of the measured particles was set to 200 MeV, a very simple color-less minimum bias matrix element has to be used. It scatters mostly valence quarks in the direction very close to the remaining proton remnants. The rest of the particle population in the event is produced through the soft remnant-remnant scattering, which is included in the eikonal model of MPI [3]. In this model, the soft MPI proceeds effectively as a gluon scattering, whose multiplicity is determined according to the Poisson distribution. It reflects the measured total cross section and the calculated hard cross section for the given settings. In fact, the generation mechanism models the outgoing gluons in the 2→4-like process, where two initial remnants change their momenta and emit two gluons, each collinear with its remnant. Herwig++ assigns the longitudinal fractional momentum as well as the transverse momentum to these gluons according to the predefined distributions. Both distributions are desired to be similar to the ones known from the perturbative scattering and thus the model fulfills the consistency requirements between hard and soft production.

It has been found that the oversimplified assignment of color charges to the outgoing gluons and proton remnants causes the rapidity gap mismodeling. Originally, the model introduced two color connections between the gluons scattered in the opposite direction assuming that such arrangement will sufficiently populate the phase space. The situation was slightly generalized by introducing the color disrupt probability parameter disconnecting these gluons completely and connecting each of them to its remnant. The new model presented at the workshop is motivated by the mentioned consistency requirement and it implements the color connections among soft gluons and remnants as in the case of QCD interaction. There are two possible arrangements of color connections. There can be a color line between the two outgoing remnants or a color line between a remnant and a gluon outgoing in the opposite direction.

Results for the $\Delta\eta^F$ distribution are now fully in agreement with the expected shape for the non-diffractive events. However, these results are obtained with the color reconnection mechanism turned off. Since this mechanism has been proven to be very useful, it will have to be improved to adopt the minimum bias measurements. One of the possibility is to introduce the space-time dependence of color reconnection algorithm which would conserve the effective amount of color connections all over the $\eta$ space.

# Measurement of the underlying event activity using charged particle jets in proton-proton collisions at 2.76 TeV

Wei Yang Wang

On behalf of the CMS Collaboration
National University of Singapore
University of Antwerp

A measurement of the underlying event in proton-proton collisions is performed using events with a leading charged particle jet produced at central pseudorapidity ($|\eta^{\text{jet}}| < 2$) and of transverse momentum ($p_T^{\text{jet}}$) in the range of 1 to 100 GeV. The analysis uses a data sample collected at the LHC by the CMS experiment at a centre-of-mass energy of 2.76 TeV, corresponding to an integrated luminosity of 0.3 nb$^{-1}$. The particle and $\Sigma p_T$ densities are studied in the transverse region, defined by the difference in azimuthal angle between the leading track jet and charged particle directions, $60° < |\Delta\phi| < 120°$. By dividing the transverse region into the minimum and maximum activity sides with respect to the highest $p_T$ jet, and studying their difference, further information on the dynamics of the UE is obtained [1].

A steep rise of the underlying event activity in the transverse region is seen with increasing leading jet $p_T$. This fast rise is followed by a saturation region above $\sim 8$ GeV, with nearly constant multiplicity and small $\Sigma p_T$ density increase. Such a distinct change of the amount of activity depending on the transverse momentum of the leading charged-particle jet is clearly seen for all the observables. The difference between the minimum and maximum activity transverse regions show an increase of activity with $p_T^{\text{jet}}$ corroborating the hypothesis of an increasing contribution from initial- and final-state radiation.

The results are compared to recent Monte Carlo event generator tunes[2, 3, 4, 5, 6, 7]. By comparing data taken at $\sqrt{s}$ = 0.9, 2.76 and 7 TeV [1, 8, 9], a strong growth with increasing centre-of-mass energy of the hadronic activity in the transverse region is also observed for the same value of the leading charged-particle jet $p_T$. All MC tunes predict the centre-of-mass energy dependence of the hadronic activity in a way that is very similar to data. These measurements are expected to play a significant role in the future development and tuning of MC models of the underlying event.

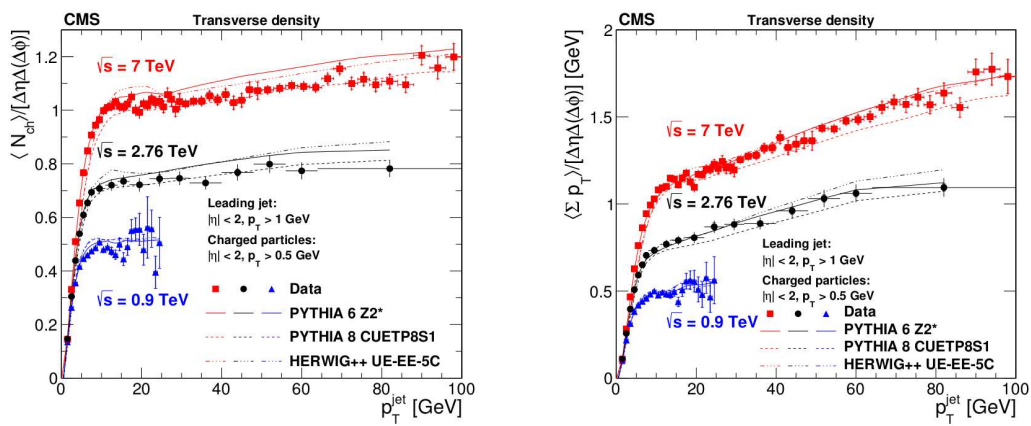

Figure 6: Comparison of UE activity at $\sqrt{s}$ = 0.9, 2.76 and 7 TeV for for particle density (a) and $\Sigma p_T$ density (b) as a function of the leading charged-particle jet $p_T$ [1, 8, 9].

Measurements of the underlying event activity has contributed to our understanding of soft dynamics and Monte Carlo generator tuning. A measurement of the underlying event activity at 13 TeV would be crucial for the verification of these tunes and provide additional discriminatory power of multi parton interactions and beam-beam remnants against initial- and final-state radiation effects on the UE activity.

# Jets in multiparticle production in and beyond geometry of proton-proton collisions at the LHC

M. Yu. Azarkin[1], M. Strikman[2]

On behalf of the CMS Collaboration
National University of Singapore
University of Antwerp

[1] Lebedev Physical Institute of RAS, National Research Nuclear University "MEPhI",
[2] 104 Davey Lab, The Pennsylvania State University, University Park, PA 16803, USA

The impact parameter (b) distribution of the minimum bias inelastic $pp$ collisions can be calculated based on the information about elastic $pp$ scattering while the distribution over $b$ of inclusive dijet production is given by the convolution of the generalized parton distributions (GPDs) of the colliding protons. The GPDs for small $x$ gluons can be extracted from the HERA data on hard exclusive processes like $\gamma p \to J/\psi p$. It was demonstrated in [1] that minimum bias and hard collisions occur, on average, at very different impact parameters – medians of $b$-distribution differ by a factor of $\approx 2$. At the same time, the b-distribution for production of dijets weakly depends on $p_T$ of the jets or their rapidity.

We use this information to analyze experimental findings by CMS on properties of jets and underlying events at different multiplicities in proton-proton interactions at 7 TeV. Observed data-MC disagreements at hight multiplicities are interpreted as an indication of increasing role of central collisions. The value of the average impact parameter of the pp collisions as a function of the soft hadron multiplicity is estimated. We find an indication that the rates of different hard processes observed by CMS and ALICE universally depend on the underlying event charged-particle multiplicity until it becomes four times higher than average. It is shown that the increase of the overlap area of colliding protons is not sufficient to explain the rate of jet production in events with charged-particle multiplicity that is more than three times higher than average. New mechanisms are necessary, like interaction of protons in rare configurations of higher than average gluon density. Such mechanisms are not included in the present Monte Carlo event generators. In any case it is interesting that the ridge in $pp$ scattering observed for multiplicities where geometry breaks and hard processes are strongly enhanced.

More detailed studies are also highly desirable. In particular, it would be preferable to study dependence on the multiplicity of the underlying event rather than on the total multiplicity. It would be also interesting to study events where $R = N_{ch}/\langle N_{ch}\rangle < 1$. In this case the ratio of double parton scattering rate to single parton scattering one should be significantly suppressed as compared to inclusive case, while for $R \sim 3$ it is expected to be enhanced as compared to the inclusive case by a factor $\sim 2$ with a further possible increase due to gluon fluctuations. It would be also interesting to investigate how the production of leading baryons (for example, production of leading neutrons with $x_F \geq 0.3$) is correlated with $R$. Indeed, it was argued [3] that the neutron yield in the proton fragmentation region drops with an increase of the centrality of the pp collisions and also decreases faster with increase of $x_F$. If so, one expects that, with increase of $R$, the neutron multiplicity for large $x_F \geq 0.3$ will diminish and it will be very strongly suppressed for $R \geq 4$.

Acknowledgments: M. Azarkin is grateful for support by the RFBR grants 12-02-91504-CERN-a, 14-02-00099, and the RAS-CERN program. M. Strikman's research was supported by DOE grant No. DE-FG02-93ER40771.

# Measurements of forward neutron and neutral pion productions with the LHCf detector

Gaku Mitsuka

University of Florence and INFN Firenze

## Energy and transverse-momentum spectra in forward region

The Large Hadron Collider forward (LHCf) experiment has been designed to measure the hadronic production cross sections of neutral particles emitted in very forward angles, including zero degrees, in proton-proton ($p$–$p$) and proton-lead ($p$–Pb) collisions at the LHC.

Figures 7 show the energy and transverse momentum ($p_\mathrm{T}$) distribution for inclusive neutral pions in very forward rapidity region has been measured with the LHCf detector in $p$–$p$ collisions at $\sqrt{s} = $ TeV7 at the LHC. The both spectra contain the recent analysis extension to the *Type-II* $\pi^0$ events, which consist of two photons entering either of two calorimeters (see the presentation at the workshop).

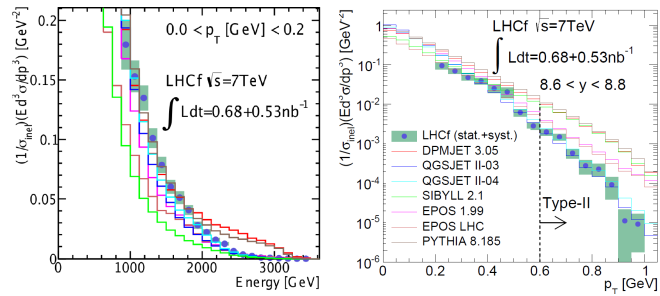

Figure 7: (Left) Energy spectrum of $\pi^0$'s at $y > 8.6$ and $p_\mathrm{T} < $ GeV0.2. (Right) $p_\mathrm{T}$ spectrum of $\pi^0$'s at $8.6 < y < 8.8$.

Figures 8 shows the energy spectra measured with the LHCf detector. The experimental results indicate that the highest neutron production rate occurs at the most forward rapidity range ($\eta > 10.76$) compared with the hadronic interaction models. QGSJET II-03 predicts a neutron production rate compatible with the experimental result at the most forward rapidity.

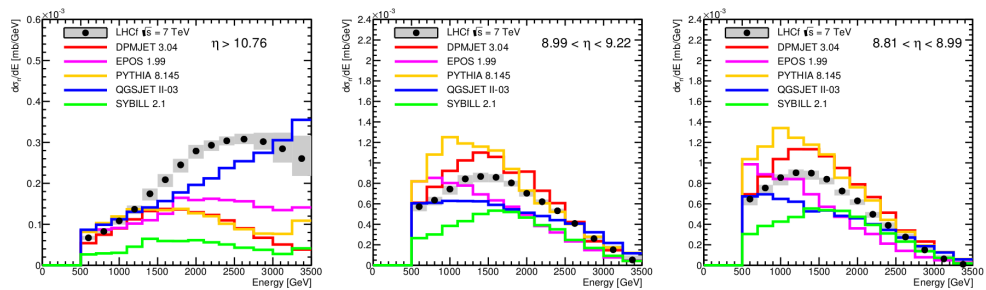

Figure 8: Energy spectra of neutrons obtained with the LHCf detector.

## Discussions

There exist several proposals (by A. Szczurek, M. Strikman, and S. Ostapchenko) to extend the LHCf capabilities to the diffractive event ragging and investigation of impact-parameter-dependent energy spectra which can be realized with a help of the ATLAS detector and a correlation between two LHCf detectors.

# Measurements of event shapes and particle production with the ATLAS detector

Róbert Astaloš

on behalf of the ATLAS Collaboration

Comenius University Bratislava, Radboud University Nijmegen

Measurements of event shapes and production properties of mesons and baryons at 7 TeV proton-proton collisions data taken by the ATLAS detector[1] in 2010 are presented.

The evolution of the event shape variables, such as the transverse thrust, thrust minor and transverse sphericity has been studied for minimum bias events as functions of the $p_T$ of the leading particle in an event[2]. With increasing $p_T^{lead}$ a shift to less spherical events is observed. This effect is more profound in case of the transverse sphericity than for the other event shape variables. All MC models considered tend to reproduce data better for higher $p_T^{lead}$ ranges. Among them the Pythia 6 tune Z1 which was tuned to Underlying Event at LHC agrees best with data. The dependence of the mean values of the event shape variables on a charged-particle transverse momentum scalar sum of the event shows an incrse followed by a decrease. For all MC models considered, this decrease occurs before that of the data (see left plot of Fig. 9 for case of the transverse sphericity mean value). Furthermore, all MC models predict less high-sphericity events than seen in the data.

The differential production cross section of the $\phi(1020)$-meson has been studied using the $\phi \to K^+K^-$ decay mode[3]. The integrated $\phi$-meson production cross section in this fiducial range ($500 < p_{T,\phi} < 1200$ MeV, $|y_\phi| < 0.8$) is measured to be $\sigma_\phi \times BR(\phi \to K^+K^-) = 570 \pm 8$ (stat) $\pm$ 68 (syst) $\pm$ 20 (lumi) $\mu$b. The cross section has also been studied as a function of the transverse momentum $p_{T,\phi}$ (right plot of Fig. 9) and rapidity $|y_\phi|$ and was found to be in best agreement with the predictions of the MC generator tunes EPOS-LHC and Pythia 6 DW.

The transverse polarization of $\Lambda$ and $\bar{\Lambda}$ hyperons produced in minimum bias events has been measured to be $P_\Lambda = -\,0.010 \pm 0.005$ (stat) $\pm 0.004$ (syst) and $P_{\bar{\Lambda}} = 0.002 \pm 0.006$ (stat) $\pm$ 0.004 (syst). In bins of $p_T$ and Feynman-$x$ variable, polarization is found to be less than 2% and is consistent with zero within estimated uncertainties as predicted by the Standard Model.

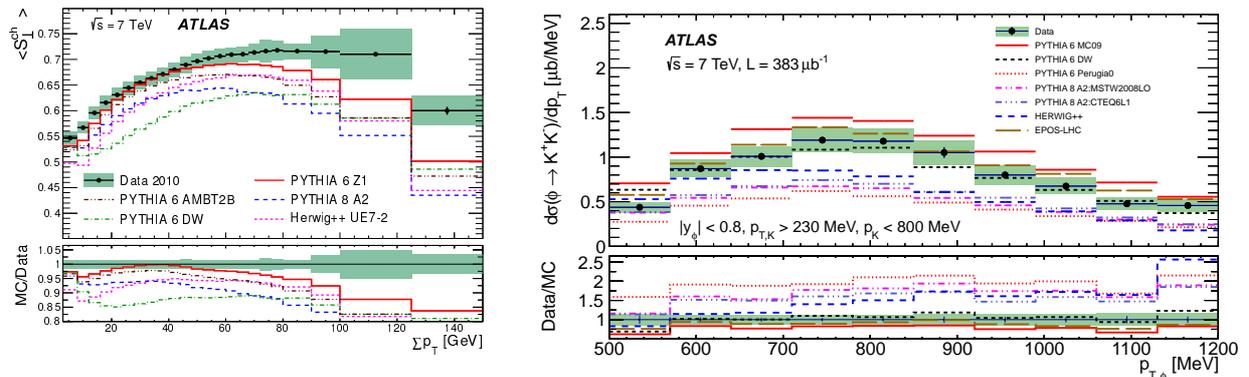

Figure 9: Mean value of the transverse sphericity as a function of charged-particle transverse momentum scalar sum of the event (left) and the $\phi(1020) \to K^+K^-$ cross section as a function of the transverse momentum $p_{T,\phi}$ (right). The error bars represent the statistical uncertainty and the green boxes represent the combined statistical and systematic uncertainty.

# References

[1] ATLAS Collaboration, JINST 3 (2008) S08003

[2] ATLAS Collaboration, Phys. Rev. D 88, 032004 (2013) [arXiv:1207.6915]

[3] ATLAS Collaboration, Eur. Phys. J. C (2014) 74:2895 [arXiv:1402.6162]

# CMS UE and MB Tunes

Rick Field

University of Florida
*For the CMS Collaboration*

## Summary

Just before the shutdown of the Tevatron, CDF collected about 12 million "Min-Bias" (MB) collisions at 300 GeV and about 54 million MB collisions at 900 GeV (the "Tevatron Energy Scan"). CDF "underlying event" (UE) measurements at 300 GeV, 900 GeV, and 1.96 TeV when combined with LHC measurements allow for a detailed study of the energy dependence of the UE. The CMS Physics Comparisons & Generator Tunes group have used the CDF data together with CMS UE measurements 7 TeV to produce the following PYTHIA tunes [1]:

- PYTHIA 6 Tune CUETP6S1-CTEQ6L: Start with Tune Z2*-lep and tune to the CDF UE data at 300 GeV, 900 GeV, and 1.96 TeV and the CMS UE data at 7 TeV. Improved version of Tune Z2*-lep using the CTEQ6L structure functions.

- PYTHIA 8 Tune CUETP8S1-CTEQ6L: Start with Tune 4C [2] and tune to the CDF UE data at 900 GeV and 1.96 TeV and the CMS UE data at 7 TeV. Improved version of Tune 4C using the CTEQ6L structure function.

- PYTHIA 8 Tune CUETP8S1-HERALOPDF: Start with Tune 4C and tune to the CDF UE data at 900 GeV and 1.96 TeV and the CMS UE data at 7 TeV. Improved version of Tune 4C using the HERAPDF1.5LO structure function.

- PYTHIA 8 Tune CUETP8M1-NNPDF2.3LO: Start with the Peter Skands Monash Tune [3] and tune to the CDF UE data at 900 GeV and 1.96 TeV and the CMS UE data at 7 TeV. Improved version of the Monash Tune using the NNPDF2.3LO structure function.

The CDF 300 GeV UE data are excluded from the PYTHIA 8 tunes becase we are unable to get a good fit to all four energies with PYTHIA 8. However, these PYTHIA 8 tunes fit the UE data at 900 GeV, 1.96 GeV, and 7 TeV very well. The PYTHIA 6 tune fits all four energies fairly well. We have extrapolated these tunes to 13 TeV. The three PYTHIA 8 tunes give almost identical predictions at 13 TeV. The PYTHIA 6 tune gives almost the same predictions at 13 TeV. Are focus has been on the energy dependence of the UE and making precise predictions at 13 TeV. We expect the early LHC UE data at 13 TeV to agree closely with the predictions of these tunes.

We have studied the productions of these tunes for the energy and particle flow in the forward region ($|\eta| > 4$). Both the original Monash tune and the CMS Tune CUETP8M1-NNPDF2.3LO agree better with the CMS-TOTEM data in the forward region than tunes with other structure functions. The low $x$ behavior of the NNPDF2.3LO gluon distribution results in more energy in the forward forward region than the other structure functions. One can increase the low $x$ gluon distribution of the the HERAPDF1.5LO structure function and improve the fits to the forward region. These modifications are in the region $x <\approx 10^{-5}$ and are below the HERA data. There is a bit of freedom to tune the low $x$ gluon distribution to improve the fits to collider data, but in doing so one has to carefully examine all the Monte-Carlo model parameters that might also affect the forward region.

## Open problems

- We do not understand why our PYTHIA 6 tune fits the CDF 300 GeV UE data better than our PYTHIA 8 tunes.

- We do not yet know if it makes sense to tune the low $x$ gluon distribution to improve the fit to collider data in the forward region. One must first carefully study all the QCD Monte-Carlo model parameters that might affect the forward region. Tuning the PDF should be the last resort.

## Outlook

I expect many interesting QCD Monte-Carlo model comparisons with the early Run 2 LHC data. It should be a very exciting year.

# ATLAS Pythia 8 tunes to 7 TeV data

Andy Buckley, on behalf of the ATLAS collaboration

School of Physics & Astronomy, University of Glasgow, UK

## Summary

The main Monte Carlo (MC) event generators in use at the LHC apply models of parton showers, hadronisation and multiple partonic interactions, augmenting fixed-order matrix element events into simulations very similar to real collider data. But the approximations inherent in these models introduce parameters into the modelling, which are fixed by "tuning" to data [2].

I presented the two most recent ATLAS collaboration [1] tunes for the Pythia 8 generator [7]: the AZ/AZNLO tunes for optimal description of the low-end of the $Z$-boson $p_T$ spectrum [4] with standalone & POWHEG [6] matrix elements; and the A14 general-purpose high-$p_T$ tunes for standalone Pythia 8 [6]. Both these tune sets were created using the Rivet [4] MC analysis system and the Professor [5] MC tuning tools.

The AZ(NLO) tunes were constructed specifically to describe the experimental $Z$ $p_T$ data better than achievable with fixed-order or analytically-resummed calculations – this is an important systematic in $W$ mass measurement. Only the Pythia primordial $k_T$ smearing parameter and initial state parton shower cutoff were tuned, the effects of these parameters being important for very low-$p_T$ recoil modelling but virtually irrelevant for other observables. In the case of AZNLO, the crucial developments were in the basic configuration: use of the POWHEG per-event `SCALUP` variable as the starting scale of the initial-state parton shower, and raising the POWHEG emission cutoff to give Pythia control over the lowest-$p_T$ shower emissions. The standalone tune describes the $Z$ $p_T$ spectrum between 1–100 GeV within 2% of the data, while the POWHEG tune is within 5% of the data for the same $p_T(Z)$ range.

Cross-checks indicate that the AZ tunes are largely viable for use as general purpose tunes, albeit with a rough MPI description. But for general-purpose MC generation a true global tune is preferred. ATLAS' standalone Pythia 8 "A14" tunes have been created for that purpose, using a large tuning space of ten parameters from the parton shower and MPI models, plus the LO hard-process $\alpha_S$. The tunes were independently constructed for the four most recent LO PDFs from the NNPDF, CTEQ, MSTW, and HERA fitting groups [9, 10, 11, 12], serving as a test of PDF impact and a consistency check on tune stability. All relevant & available observables from ATLAS' Run 1 data-taking were used, including measurements in QCD jet production, $Z$ production, and $t\bar{t}$ production processes. Cuts were placed on the tuned ranges of these observables to avoid regions where multi-jet emission dominates. Pythia 8's damped shower setup was used for $t\bar{t}$ showering; this setting was found crucial to describe $t\bar{t}$ event jet veto observables and reduce tensions between datasets.

The resulting A14 tunes show remarkable consistence between the optimal $\alpha_S$ values in initial- and final-state parton showers. Variation between PDFs are small, except in the MPI parameters, as expected. The strength of colour reconnection is also consistent between the four tunes. A set of five variation-tune pairs have also been created, using the "eigentune" method, to provide a covering set of tunes for various modelling and tuning systematics. The A14 tunes are principally intended for use in LO modelling of fairly high-$p_T$ processes, but may also be useful for showering of NLO and multi-leg events.

## Open problems

The main open problem in MC tunes for high-$p_T$ process modelling is that there is no clear path to a generator configuration suitable for use in LO, LO multi-leg, and NLO hard process showering. A well-defined ME/PS matching scheme is a crucial prerequisite for this, and this does not yet exist for the Powheg matching. Whether it is possible to create NLO tunes which can be used for example in both Powheg [6] and aMC@NLO [13], and across many process types is still an open question.

On the softer side of tuning, the A14 tune provides very good global description of high-$p_T$ phenomena but at the cost of finding a final-state shower $\alpha_S$ rather different than that found in tunes to LEP event shapes and differential jet rates only. The A14 performance on LEP observables needs to be investigated. The softest and most forward physics also continues to pose the toughest challenges: modelling of fiducial minimum bias observables and their connection to total cross-sections remains rather poor, and it seems reasonable to first aim for progress in that modelling area before aiming for consistent do-everything tunes.

Finally, the big question in soft-QCD tuning before LHC Run 2 is obviously the level of underlying event and pile-up backgrounds at 13 TeV. The A14 tunes do not address that issue but have been constructed with decoupled energy extrapolation so that adjustment can be made without any effect at 7 TeV. The spread of model predictions at 13 TeV is rather small, but surprises can always happen!

## Outlook

More ATLAS tune studies are in progress for description of minimum-bias & forward physics (for pile-up modelling), and for use with NLO/multi-leg matrix element events. Tuning of Sherpa and Herwig++ is also planned. A rapid MPI MC/data study and possible retuning is expected in response to data from the early LHC Run 2 run with a low rates of pile-up.

# Chapter:
# Double Parton Scattering

# Outlook on theory of soft and hard MPI (DPS)

Jonathan R. Gaunt

Open issues and suggestions on what can be done with Run 2 (in addition to repeating Run 1 at higher energies):

## Soft MPI, disentangling MPI and ISR/FSR

Soft activity accompanying some hard process can arise from several sources – perturbative and non-perturbative initial- or final-state radiation, and MPI. Observables that are particularly (or only) sensitive to one of these components would be a useful tool to better understand the different components of the soft activity, and improve their modelling in Monte Carlo generators. Observables that worked well at electron-positron colliders to probe final-state radiation (including hadronisation) effects were the event shape variables, and one may think that similar variables (such as beam thrust and transverse thrust) could be useful to probe ISR/FSR in proton-proton collisions. However a recent study of Glauber gluons and their connection to MPI [1] has highlighted that these observables have a strong sensitivity to soft MPI, which can formally be connected to non-cancelling Glauber exchanges between spectator partons. This sensitivity to the soft MPI can be exploited however – since the perturbative ISR/FSR component of such event shape variables is often known to high accuracy, a measurement of the event shape can be used to study the effects of the soft MPI.

What observables can then be used to disentangle the different components of soft radiation? An important study was performed in this direction recently, in the context of Z+jet or H+jet production [2]. They considered the mass of the jet, and showed that the different sources of soft radiation could be disentangled via their different dependencies on jet radius, jet $p_t$, and the partonic process. Experimental measurements of these dependencies would be valuable to get a better understanding of the soft underlying event, as would further theoretical work in identifying other potentially useful observables.

## Hard MPI (DPS), double parton distributions, correlations

Progress has been made in using constituent quark models to calculate non-perturbative properties of the double parton distributions [3]. These calculations indicate that there are quite significant correlations in longitudinal momentum and quark spin, at least in the large $x$ region in which these models should be valid. Further analyses, possibly using more sophisticated models, would be very valuable in building up a more detailed picture of the non-perturbative correlations in the proton.

Studies in which these model distributions, or other simple forms, were evolved up to higher scales using simple two-chain branching evolution, indicate that certain spin correlations (e.g. the quark-quark longitudinal spin correlation) can be non-negligible at smaller $x$ values and higher scales [3, 4]. These spin correlations can manifest themselves as correlations in the final state in DPS processes. An interesting question is how these types of correlations can be investigated experimentally, especially given that an assumption that is often made to make experimental extractions of the DPS signal is that the two scatters in DPS are uncorrelated. Double open charm production offers promise in this regard, since for this process DPS dominates over SPS and any DPS correlations are more prominent. An interesting distribution in which spin correlations may be measurable is the double differential distribution of the charm quark $p_t$s [5].

More generally, processes involving heavy quark pair production (e.g. Zcc, 4b, $J/\psi$ $J/\psi$ production at large rapidity differences) are very interesting to measure DPS, since for these the DPS/SPS ratio tends to be enhanced. Multiple production of heavy quarks in DPS takes place mainly at low $p_t$. Very valuable information could thus be obtained by identifying and measuring all heavy quarks produced in an inelastic event, with particular attention to those at low $p_t$.

Another interesting process is 4 jet production, which for DPS shows very different topologies as compared with SPS and NLL BFKL [6]. Presently missing experimental information on jets widely separated in rapidity would be relevant to identify and disentangle the different production mechanisms.

Recently much attention has been paid to the role of parton splitting mechanisms in DPS, including numerical estimations of the relative size and shape of the so called $(2 \to 2)^2$ (or 2v2) and $3 \to 4$ (or 2v1) mechanisms [7, 8]. These studies indicate that the shapes of the two contributions are rather similar, with the main effect of the $3 \to 4$ mechanism being to induce a mild dependence of the measured effective cross section on c.m. energy $s$ and scale $Q$. Experimental indications of this dependence would thus be very valuable. In particular, by covering larger ranges of the resolution scale than presently available, and reducing the experimental errors, it should be possible to reveal the QCD evolution effects.

Despite recent theoretical progress in DPS, there are still problems in a consistent definition of the double parton distributions. There are still outstanding issues related to the separation of DPS and SPS, which are linked to issues with the double parton distributions at small impact parameter. More theoretical work in this direction is needed.

Important additional information could be obtained by measuring DPS in pA interactions, where DPS are enhanced and probed in a different way as compared to pp [9]. Measuring DPS in pA in addition to pp is a way to disentangle some of the multitude of correlations possible between partons in a nucleon, and gain a more detailed insight into the nucleon structure.

# Measurement of hard double-parton interactions in $W(\to \ell\nu) + 2$ jet events at $\sqrt{s} = 7$ TeV with the ATLAS detector

Orel Gueta (on behalf of the ATLAS collaboration)

Tel Aviv University

The production of $W$ bosons in association with two jets in $pp$ collisions at a centre-of-mass energy of $\sqrt{s} = 7$ TeV has been analysed for the presence of double-parton interactions using data corresponding to an integrated luminosity of 36 pb$^{-1}$, collected with the ATLAS detector at the LHC [2]. The fraction of events arising from double-parton interactions, $f_{\rm DP}^{\rm (D)}$, has been measured through the $p_{\rm T}$ balance between the two jets. The observable, $\Delta_{\rm jets}^{\rm n}$, defined as

$$\Delta_{\rm jets}^{\rm n} = \frac{|\vec{p}_{\rm T}^{J_1} + \vec{p}_{\rm T}^{J_2}|}{|\vec{p}_{\rm T}^{J_1}| + |\vec{p}_{\rm T}^{J_2}|},$$

was chosen as it shows good discriminating power between the direct production of a $W$ boson with two jets ($W_{\rm 2j}$) and the production of a $W$ boson in association with zero jets in addition to another parton–parton scatter resulting in two jets ($W_{\rm 0j} + 2j_{\rm DPI}$). A fit is performed to the normalised, detector-level, background-corrected data distribution of $\Delta_{\rm jets}^{\rm n}$ using two normalised templates, denoted by $A$ and $B$. The result of the fit of the form $(1 - f_{\rm DP}^{\rm (D)}) \cdot A + f_{\rm DP}^{\rm (D)} \cdot B$ is shown in Figure 10. Template $A$ represents the expected contribution to the distribution of $\Delta_{\rm jets}^{\rm n}$ from $W_{\rm 2j}$ events and is selected from events generated with the ALPGEN Monte Carlo generator, interfaced with HERWIG and JIMMY (A+H+J). Template $B$ represents the expected contribution from $W_{\rm 0j} + 2j_{\rm DPI}$ events and is estimated from dijet events in data.

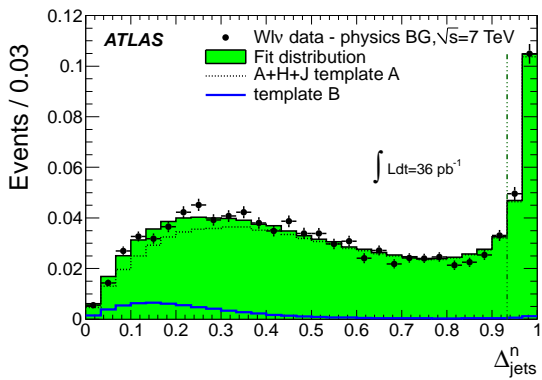

Figure 10: Distribution of $\Delta_{\rm jets}^{\rm n}$ from background-subtracted data (dots) compared to the result from the best fit for $f_{\rm DP}^{\rm (D)}$ [2]. The result of the fit is shown as the green histogram. The bins to the right of the vertical dash-dotted line were excluded from the fit. Data and the overall fit have been normalised to unity, Template $A$ (dashed line) to $1 - f_{\rm DP}^{\rm (D)}$ and Template $B$ (blue solid line) to $f_{\rm DP}^{\rm (D)}$.

For jets with transverse momentum $p_{\rm T} > 20$ GeV and rapidity $|y| < 2.8$, a central value of

$$f_{\rm DP}^{\rm (D)} = 0.08 \pm 0.01 \text{ (stat.)} \pm 0.02 \text{ (sys.)}$$

is obtained. The result for $f_{\rm DP}^{\rm (D)}$ is used to extract the parameter $\sigma_{\rm eff}$. In the formalism presented in this measurement, $\sigma_{\rm eff}$ is a universal scaling parameter, defined at the parton-level which is process and phase-space independent. It quantifies the probability for a secondary hard interaction in the same collision. The value extracted from data is

$$\sigma_{\rm eff}(7 \text{ TeV}) = 15 \pm 3 \text{ (stat.)} {}^{+5}_{-3} \text{ (sys.) mb}.$$

This value is consistent with values measured in other experiments at various centre-of-mass energies.

# References

[1] ATLAS Collaboration, New J. Phys. **15** (2013) 033038.

# DPS measurements with CMS

Diego Ciangottini and Livio Fano

Universitá degli Studi di Perugia/INFN, Italy

## Introduction

The study of DPS processes is important because it provides valuable information on the transverse distribution of partons in the proton and on the multi-parton correlations in the hadronic wave function. DPS also constitutes a background to new physics searches at the LHC.

Various measurements in pp and $p - \bar{p}$ collisions at $\sqrt{s} = 63~GeV$, 630 GeV, and 1.8 TeV are consistent with DPS contributions to multijet final states, as well as to $\gamma$ + 3-jet events at $\sqrt{s} = 1.8~TeV$ and 1.96 TeV.

The summary of DPS measurements at CMS is presented in the following sections.

## DPS in W+2jets process

The W + 2-jet process is attractive because the muonic decay of the W provides a clean tag and the large dijet production cross section increases the probability of observing DPS. Events containing a W + 2-jet final state originating from single parton scattering (SPS) constitute an irreducible background.

Events are required to have exactly one muon with $p_T > 35$GeV/c, $|\eta| < 2.1$, and missing transverse energy $MET > 30$GeV. The transverse mass is required to be greater than 50 GeV/$c^2$. Selected events are required to have exactly two jets with $p_T > 20$ GeV/c and $|\eta| < 2.0$. For DPS events, the W and the dijet system are independent of each other, while for SPS events they are highly correlated. It is thus possible to define several observables that discriminate between DPS and SPS events.

- the relative $p_T$-balance between the two jets, $\Delta^{rel}~p_T$, defined as:

$$\Delta^{rel}~p_T = \frac{|\overrightarrow{p_T}(j_1) + \overrightarrow{p_T}(j_2)|}{|\overrightarrow{p_T}(j_1)| + |\overrightarrow{p_T}(j_2)|}$$

- The azimuthal angle between the W-boson and the dijet system, $\Delta S$, defined as:

$$\Delta S = \arccos\left(\frac{\overrightarrow{p_T}(\mu, MET) \cdot \overrightarrow{p_T}(j_1, j_2)}{|\overrightarrow{p_T}(\mu, MET)| \cdot |\overrightarrow{p_T}(j_1, j_2)|}\right)$$

The fraction of DPS in W + 2-jet events is extracted with the DPS + SPS template fit to the distribution of the $\Delta^{rel}~p_T$ and $\Delta S$ observables.

The effective cross section, characterizing the effective transverse area of hard partonic interactions in collisions between protons, is calculated to be:

$$\sigma_{eff} = 20.7 \pm 0.8^{stat} \pm 6.6^{syst} mb. \tag{1}$$

The measured value of the effective cross section is consistent with the Tevatron and ATLAS results.

## DPS in 4 jets process

Scenario with four jets at different transverse momentum ($p_T$) thresholds in the final state has been measured at the CMS detector[2]. In particular, the two hardest jets are required to have $p_T$ greater than 50 GeV and they form the "hard-jet pair", while a $p_T$ threshold of 20 GeV is set for the two additional jets, comprising the "soft-jet pair". Additional jets with $p_T$ >20 GeV are vetoed. All the jets are selected in the pseudorapidity region $|\eta|$ <4.7. Single jet $p_T$ and spectra are measured and data are compared to different Monte Carlo predictions, which use a different matrix element in the perturbative QCD framework.

The $p_T$ spectrum of the hard jets is well reproduced by all the compared models at high $p_T$, while at low $p_T$, where a contribution coming from MPI is expected, the predictions start to exhibit greater differences.

The predictions including MPI need to be validated with underlying event measurements before a direct extraction of the DPS contribution can be performed via template fit. However, an alternative way to extrapolate $\sigma_{eff}$ using a MC tuning procedure has been investigated and it produces coherent results [3].

## DPS in double $J/\psi$ production

The measurement of the cross section for prompt double $J/\psi$ production in pp interactions at $\sqrt{s} = 7$TeV has been performed independently from production models [4].

A signal yield of 446±23 events for the production of two prompt $J/\psi$ mesons originating from a common vertex has been observed in a sample corresponding to an integrated luminosity of 4.73 fb$^{-1}$. The differential cross section in bins of $|\Delta y|$ is sensitive to DPS contribution. The data show evidence for excess at $|\Delta y|$ >2.6, a region that current models suggest to be exclusively populated by DPS production.

# DPS and determinations of $\sigma_{\text{eff}}$

Paolo Gunnellini, on behalf of the CMS Collaboration

DESY, Hamburg (Germany)
Universiteit Antwerpen (Belgium)

## Open problems

Measurements sensitive to double parton scattering (DPS) signals have been carried out in different final states by several experiments [1, 2, 3]. The final goal of a DPS analysis is to extract a value for the quantity $\sigma_{\text{eff}}$. Current methods used for $\sigma_{\text{eff}}$ extraction rely on the so-called "template method", which uses fits to DPS-sensitive observables performed with two different templates, one for the background contribution and one for the signal. The template method has been found to be very solid and reliable but there are some issues connected to the definition of the templates. In particular, it is difficult and sometimes even impossible to define an appropriate background template in a consistent way with different Monte Carlo event generators. This feature consequently brings larger uncertainties due to model dependence and some ambiguity in the $\sigma_{\text{eff}}$ extraction.

## Results and Outlook

A new method for $\sigma_{\text{eff}}$ determinations, called "tuning method", has been developed in order to circumvent the issues related to the definition of the templates. It is based on inclusive fits of measured DPS-sensitive observables with predictions generated with different choices of the Underlying Event (UE) parameters implemented in the simulation. The output of the fit is the tune of UE parameters which provides the best description of the data. The tune defines the corresponding value of $\sigma_{\text{eff}}$. The extraction of $\sigma_{\text{eff}}$ with the tuning method can be easily performed with the use of the RIVET [4] and the PROFESSOR [5] software. RIVET is used to generate predictions by using the corresponding plugin of the DPS-sensitive analysis. PROFESSOR is used to automatically generate random choices of the parameters, to interpolate the response of the predictions and to fit the data with the obtained predictions. The innovative part of the method is that an inclusive fit is taken without the need of dividing the contributions in background and signal.

Results have been obtained for the measurement of four-jet final states performed at the CMS experiment [6]. The tuning method has been applied by using the PYTHIA8 event generator [7] by fitting two DPS-sensitive observables, namely $\Delta S$ and $\Delta^{\text{rel}}_{\text{soft}} p_{\text{T}}$. They investigate the relative position of the jets in the transverse plane and they show sensitivity to a possible DPS signal. Values of $\sigma_{\text{eff}}$ are listed in Table 1 [8], as well as the goodness of the fits, $\chi^2/\text{NdF}$.

| Tune | $\chi^2/\text{NdF}$ | $\sigma_{\text{eff}}$ (mb) |
|---:|:---:|:---|
| P8 4C | - | 30.3 |
| CDPSP8S1-4j | 0.751 | $21.3^{+1.2}_{-1.6}$ |
| CDPSP8S2-4j | 0.428 | $19.0^{+4.7}_{-3.0}$ |

Table 1: Values of $\sigma_{\text{eff}}$ obtained for the new tunes compared to the UE-based tune 4C. The goodness of fit, evaluated as $\chi^2$ divided by the number of degrees of freedom, is also indicated for the new tunes.

Two different fits have been performed: CDPSP8S1-4j, which considers a variation of only the amount of matter overlap between the two protons and CDPSP8S2-4j which extends the tune to the whole set of UE parameters. The values of $\sigma_{\text{eff}}$ obtained for the new tunes are found to be compatible between each other around 20 mb. They are smaller than the $\sigma_{\text{eff}}$ value predicted by the UE-based tune 4C. This indicates that the DPS contribution implemented in the tune 4C needs to be increased in order to improve the description of four-jet observables.

The new approach proposed with the tuning method is very useful and efficient to extract $\sigma_{\text{eff}}$. It is very quick and easy to implement with any MC generator and any model of Multiple Parton Interaction (MPI). The tuning method requires a three-step procedure. First of all, the considered DPS-sensitive observables need to be corrected to the particle level; then the fit is performed on those. Finally the corresponding value of $\sigma_{\text{eff}}$ can be extracted and can be compared to results obtained with any other Monte Carlo event generator and physics channel. We would like to suggest some points and remarks which should be considered when using the tuning method. They are listed in the following:

- Investigation of the contribution of different matrix elements used with the same UE simulation;

- Use more than one MC event generator to study the DPS contribution needed in different models;

- Use more than one variable for the DPS determination;

- Check if the new set of parameters spoil description of more inclusive distributions.

We think that the tuning method is very reliable for determinations of $\sigma_{\text{eff}}$ and may become very important for the next LHC phase where it can be applied to measurements at a new and unexplored collision energy.

# Effects of double-parton scattering for the rapidity-distant jet production at the LHC

Antoni Szczurek[†]

Institute of Nuclear Physics, Krakow, Rzeszow University, Rzeszow, Poland

## Present status and open issues

We have discussed production of jets widely separated in rapidity. Several mechanisms have been discussed. We have concentrateed on four jet $pp \to jjjjX$ production with at least two jets with large rapidity separation in proton-proton collisions at the LHC through the mechanism of double-parton scattering (DPS). The DPS cross section was calculated in a factorized approximation. Each hard subprocess was calculated in LO collinear approximation. The LO pQCD calculations were shown to give a reasonably good descritption of CMS and ATLAS data on inclusive jet production (see [1]). It has been shown that relative contribution of DPS is growing with increasing rapidity distance between the most remote jets, center-of-mass energy in the proton-proton collision and with decreasing (mini)jet transverse momenta. We have show also result for angular azimuthal dijet correlations calculated in the framework of $k_t$-factorization approximation. Finally we have compared correlations in transverse momenta of the widely separated jets for the LHC Run II: for the DPS, $k_t$-factorization with Kimber-Martin-Ryskin unintegrated gluon distribution [3] and for illustration also for LL BFKL. We have also shown some recent predictions of the Mueller-Navelet jets in the LL and NLL BFKL framework. For the CMS configuration the DPS contribution is smaller than the dijet collinear SPS contribution even at the high rapidity distances and only slightly smaller than that for the NLL BFKL calculation. The DPS final state topology is clearly different than that for the collinear dijet SPS which may help to disentangle the two mechanisms. It is not completely understood at present what is the production mechanism of jets widely separated in rapidity as no absolute cross section was obtained so far at the LHC. We have discussed open issues and shown that the problem of separation of the different mechanisms could be addressed in Run II.

## Outlook

It is not easy to say how many jets are produced in fact for each case considered in the presentation. In the $k_t$-factorization apprach the number of jets can be bigger than two, especially when using the KMR UPDFs [2]. In the BFKL approach there can be more jets emitted in the middle of the BFKL ladder. This would require a Monte Carlo BFKL-type program (some steps in this direction have been already undertaken [4]). The number of jets in DPS can be both smaller and bigger than four (simplified LO case). We think that a study of different samples with fixed number of jets would be interesting and useful. However, detailed interpretation will require better theoretical methods and computer programs to be developed.

The DPS effects are interesting not only in the context how they contribute to distribution in rapidity distance between the jets but also per se. One could make use of correlations in jet transverse momenta, jet transverse-momentum imbalance (see [1]) and azimuthal correlations to enhance the relative contribution of DPS.

# A Light-Front approach to double parton distribution functions within constituent quark models

Matteo Rinaldi

Dipartimento di Fisica e Geologia, Università degli Studi di Perugia, and INFN, sezione di Perugia

In a recent paper of ours [1], double parton correlations have been properly investigated in the valence quark region by means of a relativistic constituent quark model through the "Light-Front Hamiltonian dynamics". As found in non relativistic (NR) previous analyses [2], also in this framework double parton correlations are present without any additional prescription and their contributions to double parton distribution functions (dPDFs), in the valence region, turn out to be sizable. In particular, in Ref. [1], dPDFs have been calculated within a fully Poincaré covariant approach which allow to overcome some inconsistencies found in the previous NR treatments, such as "the bad support problem". The most important results of Ref. [1] are the following: $i$) contributions of longitudinal momentum correlations of the interacting quarks are found to be sizable, at variance with those related to correlations in transverse momentum; $ii$) after LO-QCD evolution, in the low $x$ region accessed so far at the LHC, including for the moment being only the non singlet sector, the effect of dynamical correlations decreases in the unpolarized case, while it keeps being sizable in the spin-dependent one. In order to obtain a more complete analysis, closer to the present experimental scenario, the study of the contribution of the singlet sector to the pQCD evolution of the calculated dPDFs and the calculation of the effective cross section, entering the "pocket formula" approximation to the double parton scattering cross section, are fundamental steps which are presently under scrutiny.

# Glauber Gluons and Multiple Parton Interactions

Jonathan R. Gaunt

DESY Hamburg, Germany

Factorisation formulae are a key tool for making predictions at the LHC. Factorisation at the leading power in $\Lambda_{QCD}/Q$ (with $Q$ the hard scale) has only been rigorously proven in hadron-hadron collisions for the inclusive cross section for $p+p \to V+X$, where $V$ is colourless and $X$ can be anything, and for the cross section differential in the $p_T$ of the $V$ for this process [1].

One important difficulty in obtaining a factorisation formula is the treatment of the so-called Glauber region. A particle in the Glauber region has momentum $r$ that satisfies $|r^+r^-| \ll \mathbf{r}^2$, where $r^\pm = (r^0 \pm r^3)/\sqrt{2}$, $\mathbf{r} = (r^1, r^2)$. A Feynman graph can have a leading power contribution when some of the particles are in the Glauber region (and others are in the so-called central soft $S$, collinear $J$, or hard $H$ regions). However, unlike the $S$, $J$ and $H$ contributions, the Glauber contribution cannot readily be disentangled from the rest to achieve factorisation. In [1] it was shown that the contribution from the Glauber region cancelled for the considered observables.

'Standard' factorisation formulae containing only $S$, $J$ and $H$ contributions have been written down for other observables, and some partially proven. Here we focus on the observable of hadronic transverse energy, or $E_T$. In $p + p \to V + X$, $E_T = \sum_{i \in X} \sqrt{p_{Ti}^2 + m_i^2}$. A factorisation (or resummation) formula for this observable was written down in [2]. However, when comparing the results of this formula to the results from Monte Carlo generators incorporating multiple parton interactions (MPI), a noticeable discrepancy is seen [2]. It is interesting to ask whether the effects of MPI on this observable, missing in the standard factorisation formula, correspond to a leading power Glauber contribution, or a numerically (very) large power correction.

In [3] we addressed this question, investigating the contribution of Glauber gluons to the observable $E_T$, when $E_T \ll Q$. This was done using a model set-up in which each proton is considered to be composed only from a $q\bar{q}$ pair. We focussed on Glauber gluon exchanges between the 'spectator' partons not involved in the 'primary' $q\bar{q} \to V$ hard process. This is because it is only for these exchanges that the integral contour is trapped in the Glauber region.

It was found that the leading power Glauber contributions cancelled for one Glauber gluon exchanged between spectators, but failed to cancel at the level of two-gluon exchange (see right). The non-cancellation can be traced to a mismatch in $E_T$ value between the cut passing between the Glauber gluons (short-dashed line, MPI occurs) and those passing either side of the Glauber system (long-dashed lines, absorptive process occurs).

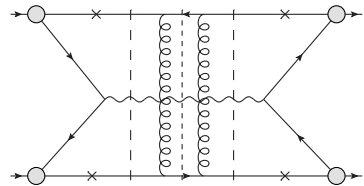

In this way a link between leading power contributions from Glauber gluons to $E_T$ and MPI effects can be established. For similar reasons one expects leading power Glauber/MPI contributions to other event shape observables in $pp$, such as beam thrust and transverse thrust. The effect of these Glauber exchanges/MPI can be suppressed by considering jet-based observables. On the other hand measurements of these event shape variables can be useful to measure MPI effects and tune Monte Carlo event generators.

# Dissecting Soft Radiation with Factorization

Wouter J. Waalewijn

Nikhef, Theory Group, Science Park 105, 1098 XG, Amsterdam, The Netherlands
ITFA, University of Amsterdam, Science Park 904, 1018 XE, Amsterdam, The Netherlands

The modeling of soft hadronic activity at the LHC relies heavily on Monte Carlo programs. We scrutinize their theoretical basis, using field theory (specifically factorization) to quantify the effect of perturbative and nonperturbative primary soft radiation. This is motivated by the ambiguity between these effects and multiple parton interactions (MPI) in underlying event measurements. We compare our predictions to Pythia 8 and Herwig++ 2.7.

Instead of looking at the regions away from hard jets, we directly study the effect of various sources of soft hadronic activity on the jet invariant mass $m_J$, which is a benchmark observable at the LHC. We show how these sources can be separated by exploiting their different dependence on the jet radius $R$, jet momentum $p_T^J$, jet rapidity $y_J$, and partonic scattering process [1].

The *normalized* jet mass spectrum in $pp \to Z+1$-jet and $pp \to H+1$-jet events is given by [2]

$$\frac{1}{\sigma}\frac{d\sigma}{dm_J^2} \simeq \int dk_S\, J_{\kappa_J}(m_J^2 - 2p_T^J k_S)\, S_\kappa(k_S, y_J, R)\,. \qquad (2)$$

Here, $\kappa = \{\kappa_a, \kappa_b, \kappa_J\}$ denotes the partonic channel, and $k_S$ is the contribution of soft radiation to the jet mass measurement. The jet function $J_{\kappa_J}$ describes energetic final-state radiation and the soft function $S_\kappa$ initial and final-state soft radiation. Eq. (2) does not include MPI.

The soft function $S_\kappa$ is sensitive to perturbative effects $S_\kappa^{\rm pert}$ and nonperturbative effects $F_\kappa$, which can be factorized $S_\kappa = S_\kappa^{\rm pert} \otimes F_\kappa$ [3]. In the tail of the jet mass spectrum, $k_S \gg \Lambda_{\rm QCD}$, $S_\kappa$ can be expanded, $S_\kappa(k_S, y_J, R) = S_\kappa^{\rm pert}(k_S - \Omega_\kappa(R), y_J, R)$. The resulting shift of the jet mass $m_J^2 = (m_J^2)^{\rm pert} + 2p_T^J \Omega_\kappa(R)$ agrees with Pythia, as shown in the left panel of Fig. 11.

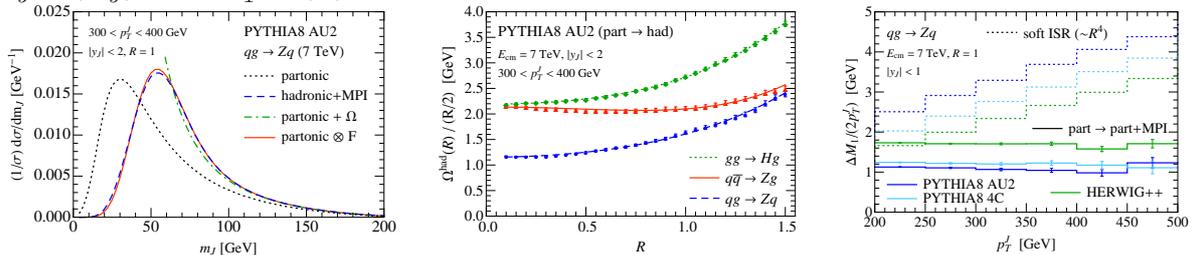

Figure 11: *Left:* The change in Pythia's jet mass spectrum from partonic to hadronic+MPI is described by a shift in the tail, and a simple convolution everywhere. *Center:* The $R$ dependence of hadronization in Pythia agrees with factorization: it fits odd powers of $R$ and for $R \ll 1$ only depends on whether the jet is initiated by a quark or gluon. *Right:* $p_T^J$ dependence of the $\sim R^4$ contributions to the jet mass in Pythia and Herwig from MPI and soft ISR for $qg \to Zq$.

The field-theoretic definition of the nonperturbative parameter $\Omega_\kappa$ allows us to show that it is independent of $y_J$, only involves odd powers of $R$, and that the linear $R$ term differs between quark and gluon jets but does not depend on the full partonic process [1]. Thus MPI cannot be part of $\Omega_\kappa(R)$ because they scale like $R^4$. Hadronization in Pythia agrees very well with our predictions for $\Omega_\kappa$, as shown for the $R$ dependence in the middle panel of Fig. 11.

The perturbative soft function contains a interference contribution from initial-state radiation (ISR) that scales like $R^4$ [1]. Current tunes do not address its potential degeneracy with MPI. As the right panel in Fig. 11 shows, these $R^4$ contributions differ much more between tunes than their sum, and can be separated by exploiting their different $p_T^J$ and channel dependence.

# On the longitudinal structure double parton distributions

Federico Alberto Ceccopieri

IFPA, Université de Liège, 4000 Liège, Belgium

## Open problems

At high energies, multiple parton scattering may occur within the same hadronic collision. As long as the scattering processes are characterised by large momentum transfers we may try to generalise, as a first approximation, the concepts of hard scattering factorisation derived in the single particle scattering context. In the case of double parton scattering, under the simplifying assumptions of two uncorrelated hard scatterings, measurements have provided so far informations only on $\sigma_{eff}$, the dimensionful parameter which controls the overall relative magnitudo of the double parton scattering contribution, with no strong indications of its energy, hardness and flavour dependence. Among channels which have been proposed to study the multiple parton component of the scattering, we mention here the associated production of a particle and a electroweak boson, along the lines described in Ref. [1]. This processes, triggering on particles rather than on jets, can access the low but sill perturbative transverse momentum region of the particle production spectrum, which is expected to be populated by particles generated in additional partonic interactions. Moreover, the possibility to control initial state radiation by properly restricting the transverse momentum of the electroweak boson can give additional handle in the characterisation of the double or multiple parton scattering component. On the theoretical side, double parton scattering gives access to parton correlations in the hadron structure not accessible in single parton scattering and encoded in novel distributions, double parton distributions. The factorisation of the relevant cross sections is presently under scrutiny [2, 2] and the relevant issue of the proper scale evolution of these distributions has been already pointed out. Its solution is mandatory if the whole double parton scattering formalism has to be properly validated in different processes at different energies. If the dependence on relative interparton distance is integrated over, one obtains longitudinal double parton distributions evolution equations [4, 5]. These distributions however do not appear directly into double parton scattering cross sections and their evolution is characterised by an additional term, responsable for perturbative longitudinal correlation between the interacting partons, which poses a problem of double counting with single parton scattering cross sections. Moreover, when viewed in position space, such term accumulates at zero transverse distance exhibiting a singular behaviour in this limit [2]. In order to solve these problems, it is therefore necessary to introduce distributions unintegrated over relative transverse parton momenta, addressed as generalised parton distributions, $_2$GPD$s$, which are directly connected to the value of $\sigma_{eff}^{-1}$ [6] without factorising assumptions. On the other hand, the inhomogeneous term discussed above is essential to preserve the expected sum rules for longitudinal double parton distributions [7]. It appears therefore that the road towards a consistent treatment of QCD evolution effects of (generalised) double parton distributions is quite narrow as it must reconcile all these requirements at once.

## Outlook

We have outlined measurements potentially sensitive to double partonic interactions and some theoretical aspects of the present multiple scattering theory which requires further improvements.

# Electroweak boson production in double parton scattering

Emilia Lewandowska and Krzysztof Golec-Biernat

Institute of Nuclear Physics Polish Academy of Sciences
Krakow, Poland

## Open problems

We study the $W^+W^-$ and $Z^0Z^0$ electroweak boson production in double parton scattering (DPS) using QCD evolution equations for double parton distribution functions (DPDFs) [1]. In particular, we analyse the impact of splitting terms in the evolution equations on the DPS cross sections. Unlike the standard terms, the splitting terms are not suppressed for large values of the relative momentum of two partons initiating the DPS, thus they play an important role for the DPS cross sections.

## Outlook

In our work, we have analyzed the DPS processes in the collinear approximation, using QCD evolution equations for the DPDFs. We have concentrated on the significance of the splitting terms in these equations for the DPS processes with a large hard scale $Q \sim 100$ GeV. For the illustration, we considered the $W^+W^-$ and $Z^0Z^0$ boson production at the LHC center-of-mass energy 14 TeV. To compute the DPS cross sections, we have specified the dependence of the DPDFs on the relative momentum $\mathbf{q}$. The DPS cross sections can be written as the sum of two components

$$\sigma_{AB} = \sigma_{AB}^{(11)} + \sigma_{AB}^{(12+21)} \tag{3}$$

in which

$$\begin{aligned}\sigma_{AB}^{(ij)} &= \frac{N}{2} \sum_{f_i, f_i'} \int dx_1 dx_2\, dz_1 dz_2 \int \frac{d^2\mathbf{q}}{(2\pi)^2}\, \theta(Q - |\mathbf{q}|) \\ &\times\, D_{f_1 f_2}^{(i)}(x_1, x_2, Q, \mathbf{q})\, \hat{\sigma}_{f_1 f_1'}^A(Q)\, \hat{\sigma}_{f_2 f_2'}^B(Q)\, D_{f_1' f_2'}^{(j)}(z_1, z_2, Q, -\mathbf{q}).\end{aligned} \tag{4}$$

The splitting component, $\sigma_{AB}^{(12+21)}$, is not strongly suppressed at large values of $|\mathbf{q}|$, like the standard component, $\sigma_{AB}^{(11)}$, because it originates from the splitting of a pointlike parton. Based on the constructed numerical program, which solves the evolution equations for DPDFs, we analyzed the single splitting contribution to the DPS cross sections for the electroweak boson production. We quantified the relevance of the single splitting contribution in such a case in terms of the effective cross section. We also discussed the correlations in rapidity for the produced $W^\pm$ bosons pointing out that the single splitting contribution distorts the standard correlation obtained with the factorized DPS cross section.

# Correlations in $J/\psi$ pair production as SPS versus DPS discriminators

Sergey Baranov

P.N.Lebedev Institute of Physics
Russian Academy of Sciences

## Motivation and theoretical framework

We focus on the problem of disentangling the single (SPS) and double (DPS) parton scattering modes in the production of $J/\psi$ pairs at the LHC conditions. We carefully examine the production properties in the different kinematical domains paying attention to the different contributing processes and compare the shapes of the differential cross sections.

Our approach is based on perturbative QCD, nonrelativistic bound state formalism, and the $k_t$-factorization ansatz in the parton model. On the SPS side, we consider the leading-order $\mathcal{O}(\alpha_s^4)$ contribution (31 box diagrams calculated in [1]) and the subleading $\mathcal{O}(\alpha_s^6)$ contribution from pseudo-diffractive gluon-gluon scattering represented by one gluon exchange and two gluon exchange mechanisms. The latter can mimic the DPS mechanism having similar kinematics. On the DPS side, we consider the prompt $J/\psi$ production including the direct $g+g \to J/\psi + g$ contribution and radiative $\chi_c$ decays. All necessary details of calculations can be found in [3].

## Results and conclusions

Naive expectations that the SPS mechanism should result in back-to-back event configuration received no support from calculations. Including the initial state radiation effects (either in the form of $k_t$-dependent gluon densities [1] or by means of simulating parton showers in a phenomenological way [2]) washes out the azimuthal correlations, thus making the SPS and DPS samples very similar to each other. The correlations get restored at higher $p_t$, but then the production rates fall dramatically, and so, the practical discrimination of the production mechanisms remains problematic.

Selecting the events with large rapidity difference between $J/\psi$ mesons looks more promising. As it was expected, the DPS production leads to broad and relatively flat distribution in $\Delta y$. The leading order SPS contribution is localized inside the interval $|\Delta y| \leq 2$ (and continues to fall down steeply with increasing $|\Delta y|$), while the higher order pseudo-diffractive contributions extending beyond these limits are heavily suppressed by the color algebra (see [3] for details) and do not constitute significant background for the DPS production.

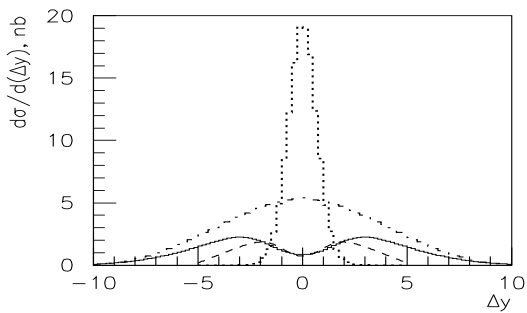

Distributions over the rapidity difference between $J/\psi$ mesons. Dotted curve, SPS 'box' contribution; dashed curve, one-gluon exchange contribution multiplied by 1000; solid curve, two-gluon exchange contribution multiplied by 25; dash-dotted curve, DPS production.

# Double open charm production at the LHC within single- and double-parton scattering

Rafał Maciuła

Institute of Nuclear Physics Polish Academy of Sciences

## Outlook

We discuss production of two pairs of $c\bar{c}$ within a simple formalism of double-parton scattering (DPS). Surprisingly large cross sections, comparable to single-parton scattering (SPS) contribution, are predicted for LHC energies. We compare results of exact calculations of single-parton scattering (SPS) and double-parton scattering (DPS) for production of $c\bar{c}c\bar{c}$ and for $D-D$ meson-meson correlations [1]. Each step of DPS is calculated within $k_t$-factorization approach, i.e. effectively including higher order corrections. The SPS calculations are performed in collinear approximation with exact matrix element for $gg \to c\bar{c}c\bar{c}$ subprocess as well as with approximate matrix elements in high-energy approximation. We compare our predictions for double charm production (DD meson-meson pairs) with recent results of the LHCb collaboration for azimuthal angle $\phi_{DD}$, dimeson invariant mass $M_{DD}$ and rapidity distance between mesons $Y_{DD}$. The predicted shapes are similar to the measured ones, however, some strength seems to be lacking. Our calculations clearly confirm the dominance of DPS in the production of events with double charm. At the LHC energy $\sqrt{s} = 7$ TeV one observes unprecedent dominance of double parton scattering. At the nominal LHC energy $\sqrt{s} = 14$ TeV the dominance would be even larger. This opens a possibility to study the DPS effects in the production of charmed mesons.

## Open problems

We have calculated several distributions and compared our results with experimental data of the LHCb collaboration. Our DPS mechanism gives a reasonable interpretation of the measured distribution. Some strength is still missing. This rather cannot be explained by single-ladder-splitting contributions to double-parton scattering production (see e.g. Ref. [2]) which are not taken into account in this analysis. Of course, given large theoretical uncertainties related to $\sigma_{eff}$ or with scales and charm quark mass the problem may not be easy to identify and solve. In order to make more definite conclusions, one should rather concentrate on violation of simple factorized Ansatz for DPS. Explicit treatment of the full-form correlation factor $F(b; x_1, x_2, \mu_1^2, \mu_2^2)$ in double-parton distributions is expected to result in reduced value of $\sigma_{eff}$ with increasing scales $\mu_1^2$, $\mu_2^2$ [3]. Another idea is to include spin correlations [4]. For conventional DPS mechsanism these effects are expected to be rather small after evolution, especially when both partons are gluons. However, it might be interesting to do a more detailed study of the spin effects, also including their effect in the single-ladder-splitting mechanisms. This is particularly interesting in light of the experimentally observed azimuthal correlations between two $D^0$ mesons produced in proton-proton collisions. Still open but very important question is the DPS contribution to inclusive single particle spectra. We emphasized significant contribution of DPS mechanism to inclusive charmed meson spectra measured recently by ALICE, ATLAS and LHCb [1]. Clear statement in this subject should be worked out by the MPI community.

# Conventional versus single-ladder-splitting contribution to double gluon-gluon initiated processes

Jonathan R. Gaunt[1], Rafał Maciuła[2] and Antoni Szczurek[2,3]

[1]DESY Hamburg, Germany, [2]Cracow, INP, Poland & [3]Rzeszów U., Poland

It is known that there are (at least) two different types of contribution to the double parton scattering (DPS) cross section. One of these is the well-known 'conventional'/2v2 contribution in which two separate ladders emerge from both protons and interact in two separate hard interactions. The other type of process is the 'perturbative ladder splitting'/2v1 contribution, which is similar to the 2v2 process except that one proton initially provides one ladder, which later perturbatively splits into two. In [1] we performed a study of the ladder splitting effects in several DPS processes in which gluon-gluon fusion is the dominant mechanism for both subprocesses – namely double quarkonium, $HH$, and $c\bar{c}c\bar{c}$ production (see also [2] for a similar study in the context of four-jet, $\gamma + 3j$, $W + jj$ and $W^+W^-$ production).

In order to calculate the 2v2 and 2v1 DPS cross sections one needs values for both the 'independent ladder pair' (ILP) and 'ladder splitting' (LS) double PDFs (dPDFs). Schematically, $\sigma(2v2) \propto$ ILP dPDF $\otimes$ ILP dPDF $\times 1/\sigma_{eff,2v2}$ and $\sigma(2v1) \propto$ ILP dPDF $\otimes$ LS dPDF $\times 1/\sigma_{eff,2v1}$, with $1/\sigma_{eff,2v1}$ and $1/\sigma_{eff,2v2}$ geometrical prefactors. We took a simple product of MSTW2008LO PDFs to be the ILP dPDF. The LS dPDFs were calculated using MSTW2008LO PDFs for the initial single ladder, and permitting splitting starting at $Q_0 = 1$ GeV. Following simple model considerations we took $\sigma_{eff,2v2}/\sigma_{eff,2v1} = 2$ and $\sigma_{eff,2v2} = 30$ mb.

In the first part of the study the LS and ILP dPDFs were compared directly at scales $\mu_1, \mu_2 = 10$ GeV. The ratio of LS/ILP was found to be rather flat for various parton combinations at small $x_1, x_2 \lesssim 10^{-3}$. The reason for this is that the ladder splitting in the LS dPDF tends to occur very early – this means that the LS and ILP dPDFs to have the same 'two-ladder' evolution over most of their evolution range, which causes their low $x$ shapes to converge.

Next 2v1 and 2v2 cross sections were compared for the double quarkonium (& $HH$) process. Values of $R = \sigma(2v1)/\sigma(2v2)$ between $0.4 - 1.3$ were obtained varying $\sqrt{s}$ of the $pp$ collision between $0.2 - 13$ TeV, and $M$ of produced heavy particles between $3 - 126$ GeV. Increasing $\sqrt{s}$ decreased $R$, whilst increases in $M$ increased it. Plots of $R$ against the rapidities of the produced particles for fixed $\sqrt{s}, M$ were also produced, and were found to be rather flat as a result of the similar $x$ shapes of the underlying dPDFs.

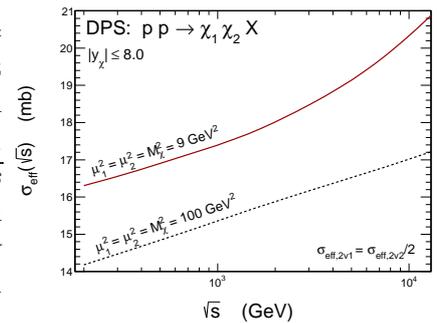

Another quantity of interest to plot is the empirical $\sigma_{eff} = \sigma_{A,B}(DPS)/[\sigma_A(SPS)\sigma_B(SPS)]$ (SPS = single scattering) that is often extracted in experiments. If only the 2v2 mechanism were present this quantity would be approximately constant, whereas with the two components it increases slowly with $\sqrt{s}$ and decreases slowly with scale (a plot for double quarkonium is given to the right). The variation of $\sigma_{eff}$ with scale is a key prediction of the two-component DPS framework, and it would be interesting to test this if possible at the LHC experiments.

Finally the 2v1 and 2v2 cross sections were compared for the $c\bar{c}c\bar{c}$ process. The results obtained were broadly similar to those for the double quarkonium process. For this process one can plot more differential distributions, such as the $p_T$ and rapidity of the charm particles. The 2v1 and 2v2 contributions to these distributions were found to have very similar shapes, making it difficult to disentangle the 2v1 and 2v2 contributions via these quantities.

# Towards correlations in double parton distributions

Alexander Snigirev[a], Nina Snigireva[b], Gennady Zinovjev[c]

[a]Skobeltsyn Institute of Nuclear Physics,
Lomonosov Moscow State University, 119991, Moscow, Russia

[b]School of Mathematical Sciences,
University College Dublin, Belfield, Dublin 4, Ireland

[c]Bogolyubov Institute for Theoretical Physics,
National Academy of Sciences of Ukraine, 03680 Kiev-143, Ukraine

The basic properties of double parton distribution functions (dPDFs) are reviewed in the framework of perturbative QCD in the collinear approximation. The initial state of double parton scattering (DPS) is coded by dPDFs which quantify the joint distribution of two partons in a hadron, depending on their quantum numbers, longitudinal momentum fractions and the relative transverse distance between them. The dPDFs and the corresponding evolution equations are well-known only integrated over the parton pair transverse separation in the collinear approximation. The particular solutions of these non-homogeneous equations may be presented in form of a convolution of single distributions (which coincides with jet calculus rules proposed originally for the fragmentation functions). They contribute to the inclusive cross section of DPS with a larger weight (different effective cross section) being compared to the solutions of homogeneous equations. The latter solutions are usually approximated by factorized form if initial nonperturbative correlations are absent.

Using the explicit form of the solutions of evolution equations in the Mellin representation we demonstrate [1] analytically that the dPDFs "forget" the initial correlation conditions (unknown *a priori*) at not parametrically small longitudinal momentum fractions $x_1$ and $x_2$, and the correlations perturbatively calculated survive only in the limit of large enough hard scales. Such a dominance is the mathematical consequence of the relation between the maximum eigenvalues $\Lambda(n)$ in the moments representation, $\Lambda(n_1 + n_2) > \Lambda(n_1) + \Lambda(n_2)$, in QCD at large $n_1$ and $n_2$ (finite $x_1$ and $x_2$). It is independent of the strength of the initial correlation conditions at all. The numerical estimates (done by integrating directly the evolution equations) testify in favour of conclusion also that the perturbative correlation effects are significant in the kinematical region accessible in experimental measurements at energies of Tevatron and LHC.

Unlike naively accepted expectations, the QCD dynamical perturbative correlations result effectively in the dependence of the experimentally extracted effective cross section of DPS on the resolution scale. The measurements covering a larger range of the resolution scale variation might reveal the evolution effects more distinctly in accordance with the asymptotic QCD behaviour.

# DPS and the Higgs signal strength at the LHC

Mieczyslaw Witold Krasny

LPNHE, Universités Paris VI et VII and CNRS–IN2P3, Paris, France

## Purpose

The principal purpose of this talk was: (1) to bring to an attention, and (2) to stimulate discussions, within the Double Parton Scattering (DPS) expert community, on the loopholes in our present theoretical and experimental understanding of the vector bosons pair (VV) production at the LHC. In the region in which the transverse momenta of both bosons are small, the kinematical characteristics of leptons produced by the conventional sources mimic very closely those of the Higgs boson decays. To establish beyond any doubt the evidence of the SM Higgs boson and to determine the strength of its coupling to vector bosons additional calculations are needed. They must merge, so far distinct, efforts of the "QCD-boxes" and the "DPS" communities to address the interplay between the gluon-gluon scattering box diagrams and the full set of the DPS processes, with a special attention to the DPS collinear singularity region.

## DDYP and the Higgs searches background

The specific type of the DPS: the Double Drell-Yan Process (DDYP), and its contribution to the vector boson pair production at the LHC was discussed in [1], where the WINHAC generator was adopted to study the DDYP contribution within the approximation in which the amplitude interferences, colour, spin and flavour correlations, correlations between the longitudinal momenta of participants (beyond the momentum conservation constraints) were neglected. Already within such an approximation the DDYP contribution was found to mimic rather closely the Higgs boson decay signals, both in the $4l$ and $2l2\nu$ channels, provided that the $\sigma_{eff}$ for the scattering of the same flavour quark-antiquark pairs (for which no experimental constraint exists) is significantly smaller than that for the quark-gluon and gluon-gluon pairs.

Several possible sources of enhancements of the DDYP probability were discussed in this contribution: (1) the reduced transverse plane correlation length for the same flavour, opposite charge quark-antiquark pairs; (2) the flavour correlation effects in proton excitations satisfying the local charge and flavour conservation; (3) the spin effects assuring the local spin compensation within the proton; and (4) the alternative mechanism of the electro-weak symmetry breaking, enhancing the VV cross-sections.

## Outlook

The DDYP contribution to the Higgs background was assumed by the ATLAS and CMS collaboration to be negligible. This is indeed the case for the simpleminded model lacking flavour, longitudinal momentum, transverse momentum and spin correlations between colliding partons. In our view the above conclusion is premature and needs further investigation of several reasons of the enhancement of the DDYP cross-sections were presented in this talk. The main purpose of the talk was to advocate a need for a new theoretical calculations beyond the leading-order and leading-twist approximations and, perhaps more importantly, the necessity of a more open investigation of the properties of observed excesses of VV events at the LHC – beyond the SM Higgs decay context.

# The study of QCD in high pile-up environments with Higgs boson and Drell-Yan production

H. Van Haevermaet

Deutsches Elektronen Synchrotron, D 22603 Hamburg
Universiteit Antwerpen, Elementaire Deeltjes Fysica, B 2020 Antwerpen

## Open problems

The study of QCD is typically done using measurements of processes in single proton-proton collisions. With its clean final state, a particular interesting one is the Drell-Yan process, which can be used to precisely measure the quark structure functions, parton showers, and underlying event properties. The properties and structure functions of gluons however, have to be determined indirectly. With the recent discovery [1] of the Higgs boson, this picture changes. In the heavy top limit, the Higgs boson directly couples to gluons and can be produced with the gluon fusion process through a colour singlet current [2]. The access to this new process opens up a whole new interesting area of possible QCD measurements, where one can directly study gluon induced effects [3]. However, due to the small cross sections, current accelerators like the LHC at CERN have to operate at very high beam intensities to collect sufficient statistics. This allows one to measure differential distributions, but it also creates a condition (called pile-up) in which on average much more than one proton-proton collision happens per bunch crossing. This implies that the phase space will be completely filled with extra hadronic activity coming from the pile-up events, which makes a study of QCD in the Higgs or Drell-Yan channels extremely difficult.

## Outlook

In this study we presented a novel method that compares Higgs and Drell-Yan production in the same invariant mass and rapidity range to perform a direct measurement of gluon versus quark induced processes. In particular we studied the underlying event activity in Higgs and Drell-Yan processes by looking at the charged particle multiplicity and scalar $\Sigma p_T$ in the transverse region. We constructed the subtracted underlying event observable, and comparing pile-up = 0, 20, and 50 scenarios we can conclude that it is stable in high pile-up environments. Thus, by comparing Higgs to Drell-Yan production we can subtract the pile-up contributions, and directly measure the difference in gluon versus quark induced initial state radiation effects, such that one can still access (small-$p_T$) QCD physics in high pile-up environments. Furthermore, this method allows for a completely detector independent pile-up subtraction that is not limited to the small experimental tracker ranges where vertex identification can be used. Hence it does not introduce any extra bias or uncertainty due to such complicated corrections, and even allows to measure initial state radiation in more forward regions. Finally, we are in the process of applying the same kind of strategy to vector boson + jet topologies to study the high-$p_T$ region of QCD.

Chapter:
# MPI and small-$x$ and diffraction

# Outlook on MPI and small-$x$ and diffraction

O. Kepka, P. van Mechelen, W. Schäfer, A. Staśto

In models of multiparticle production at high energies based on Pomeron exchanges, multiple interactions and diffractive processes are part and parcel of the formalism. While the former are related to diagrams with multiple cut Pomerons, diffractive processes, or rapidity gap fluctuations, are described by the exchange of uncut Pomerons.

The perturbative QCD gluon ladder can be regarded as representing a cut Pomeron. In multipomeron exchange models, cut and uncut Pomerons enter with such combinatorial weights (given by the AGK cutting rules), that the single particle inclusive cross section at central rapidities is just given by the contribution of a single cut Pomeron. All effects of multiple interactions/rescatterings cancel in this observable.

A similar cancellation of spectator interactions is also part of the standard hard scattering factorization theorems for inclusive hard processes. At small-$x$, there is not yet any coherent formalism available that would allow us to go to less inclusive observables, like for example multiple hard processes, hard processes as a function of the underlying event multiplicity etc.

The search for novel QCD dynamics at low $x$ is centered around effects related to the transverse momentum of partons entering the hard scattering. Various observables have been proposed, such as the azimuthal balance of jets, the likelihood to produce additional jets within or outside the rapidity range covered by the leading dijet system, etc. Investigations using the LHC run 1 data have shown an intricate interplay between low $x$ effects and multiple partonic interactions. This points to the need for a unified theoretical approach including low $x$, transverse momentum dependent parton densities and multiple partonic interactions.

Multiple interactions are also important for the understanding of diffractive events: here the interaction of spectator partons in a diffractive process would fill the rapidity gaps. One needs to know the probability that no additional interaction occurs – the so-called gap survival probability. Lepton pair production with rapidity gaps is particularly interesting in this regard. Here the lepton-pair production via photon-photon fusion at "Born-level" is calculable from QED and proton structure functions. A deviation of the measurement from such a calculation may determine the probability that protons are transmitted through the interaction region intact. Such an analysis appears to be possible in the upcoming LHC run.

Another way to investigate the relation between rapidity gap survival and multiple partonic interactions is to measure the fraction of events with a (forward) rapidity gap w.r.t. inclusive events, as a function of (central) jet transverse momentum. This is reminiscent of the study of the underlying event by looking at the activity transverse to a leading particle or jet as function of the particle or jet $p_{\rm T}$. For a fixed gap size, the fraction of events with a gap will decrease as function of $p_{\rm T}$ and it would be interesting to investigate whether phenomenological models for multiple partonic interaction can describe this behaviour well.

The ATLAS and CMS collaborations will upgrade their detectors with forward proton taggers during the LHC run 2. Several upgrade stages are foreseen. This will allow to tag diffractive events and measure the proton kinematics, even in the presence of pile-up and is thus a very attractive development for the study of hard diffractive processes with a small cross section.

# Production of leading charged particles and leading charged-particle jets at small transverse momenta with the CMS detector

Albert Knutsson on behalf of the CMS collaboration

Antwerpen University, Belgium

Normalized cross-sections of leading charged particles and charged-particle jets are measured in proton-proton low pileup data at $\sqrt{s} = 8$ TeV, where "leading" refers to the single particle or single jet with the highest transverse momentum in each event, in data collected with the CMS detector at the CERN LHC. The particles and jets are measured, respectively, in pseudorapidity ranges of $|\eta| < 2.4$ and $|\eta| < 1.9$ for transverse momenta $p_\mathrm{T} > 0.8$ GeV and $p_\mathrm{T} > 1.0$ GeV. The measured yields integrated above a given minimum transverse momentum, $p_\mathrm{T,min}$, provide information of how parton-parton cross section calculations can be improved when used in low-$p_\mathrm{T}$ nonperturbative regimes. Comparison of data with predictions from models as implemented in various event generators shows that the measured observable is sensitive to key aspects in the transition between soft and hard QCD physics. The results are shown in Figure 12.

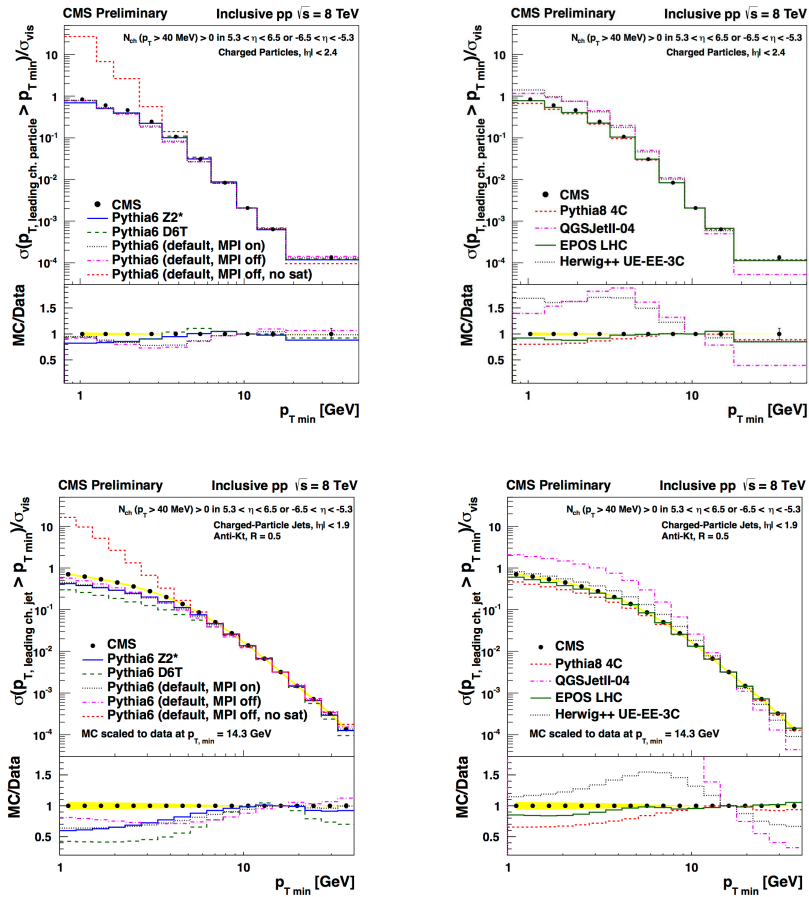

Figure 12: Normalised integrated $p_T$-distribution of the leading charged particle in $|\eta| < 2.4$ (upper) and the leading charged jet in $|\eta| < 1.9$ (lower). The data are compared to different predictions from various PYTHIA6 tunes (left) and various Monte Carlo event generators (right). The error bars indicate the statistical uncertainty and the shaded area the systematic uncertainty. The Monte Carlo curves for the leading particles and the leading jets are normalised to the measured value of $\sigma(p_{\mathrm{T\,leading\,ch.particle}} > 9.0\ \mathrm{GeV})/\sigma_{vis}$ and $\sigma(p_{\mathrm{T\,leading\,ch.particle}} > 14.3\ \mathrm{GeV})/\sigma_{vis}$, respectively, where $\sigma_{vis}$ is defined by the number of events with at least one charged particle with $|\eta| < 2.4$ and $p_t > 0.4$ GeV.

# Studies of azimuthal decorrelation of dijets with a jet veto with ATLAS

Vojtěch Pleskot

Institute of Particle and Nuclear Physics
Charles University in Prague

## Open problems

The fixed-order next-to-leading QCD predictions provide good description of the inclusive jet cross sections up to the highest reached jet energies [1, 2]. For certain observables and in some regions of phasespace higher order contributions are important and must be resummed to all orders. It is the case of dijet events with large rapidity separation of the two jets, $\Delta y$, and/or when a veto is applied on additional jets with $p_T > Q_0$ produced in the rapidity interval between them. $Q_0$ is called a veto scale. There are two available resummation approaches, BFKL and DGLAP. BFKL resums the higher orders in terms of $\ln(1/x)$. DGLAP does it in terms of $\ln(Q^2)$. The goal of the presented measurement [3] is to test theoretical predictions inspired by the above two approaches. The DGLAP inspired predictions are POWHEG+Pythia 8 and POWHEG+Herwig 6.5. POWHEG [4, 5, 6] provides full NLO dijet calculation. It is interfaced to Pythia 8 [7] or Herwig 6.5 [8] to provide the parton shower. The predictions based on BFKL are HEJ [9, 10] and HEJ+Ariadne. HEJ resums leading-logarithms relevant in the Mueller-Navelet limit. ARIADNE [11] provides parton showers based on the colour-dipole cascade model.

Several observables are measured with the data sample recorded by ATLAS in 2010 and 2011 in proton-proton collisions at 7 TeV centre-of-mass energy. The 2010 data sample corresponds to $36.1 \pm 1.3$ pb$^{-1}$ and was taken under low pile-up conditions. $Q_0$ is set to 20 GeV. Jets with $|y| < 4.4$ are accepted. The 2011 sample has $4.5 \pm 0.1$ fb$^{-1}$ and it is more affected by pile-up. $Q_0$ is set to 30 GeV and jets are accepted if $|y| < 2.4$. An additional requirement of $\Delta y > 1$ was used on the 2011 sample to enhance the events of interest.

Dijet events were selected if the (sub)leading jet $p_T$ was higher than 60 (50) GeV. The events were classified into two categories: the inclusive and gap dijet events. An event contributes to the gap sample if there is no jet with $p_T > Q_0$ in the rapidity interval between the two leading-$p_T$ jets.

An interesting observable is the gap fraction, the ratio of the dijet cross sections measured on the gap and the inclusive samples. It is exponentially suppressed as a function of $\Delta y$ and $\log(\bar{p}_T/Q_0)$ where $\bar{p}_T$ is the average $p_T$ of the two leading jets. There is a plateau at high $\Delta y$ and high $\bar{p}_T$ that can be explained as an effect of steeply falling parton distribution functions and/or as a consequence of colour-singlet exchange processes. Complementary observable is the mean multiplicity of jets with $p_T > Q_0$ in the rapidity interval between the two jets. POWHEG+Pythia 8 and HEJ+ARIADNE provide good description of both variables. Herwig 6.5 worsens the POWHEG prediction with respect to Pythia 8.

The studied observable with the highest discrimination power between different predictions is the ratio of the first two angular moments: $\langle\cos(2\Delta\phi)\rangle/\langle\cos(\pi-\Delta\phi)\rangle$ where $\Delta\phi$ is the azimuthal separation of the two jets. This observable is especially sensitive to the theoretical differences in the regions of low $\bar{p}_T$ and high $\Delta y$. HEJ+ARIADNE provides best description of the ratio on the inclusive dijet sample; its performance is a bit worsened on the gap sample.

The double differential dijet cross section $d^2\sigma/d\Delta\phi d\Delta y$ is not well described by the predictions using HEJ. It is quite well described by POWHEG+Pythia 8.

## Outlook

No theoretical prediction describes data on the full phasespace and for all observables. Data can constrain parton-shower models because the experimental uncertainty is generally lower than the spread of the different predictions. Theoretical predictions should be improved to decide about the presence or absence of BFKL effects or colour-singlet exchange processes.

# Dijet production at HERA and tests of QCD factorisation

Lidia Goerlich
on behalf of the H1 and ZEUS Collaborations

Institute of Nuclear Physics
Polish Academy of Sciences

## Exclusive dijet production in diffractive DIS

The exclusive production of dijets in diffractive deep inelastic lepton-proton scattering has been measured by the ZEUS Collaboration for the first time at HERA [1]. Jets have been reconstructed in the virtual photon-Pomeron centre-of-mass system frame using the exclusive $k_T$ algorithm. The cross section for the exclusive production of dijets is given as a function of the azimuthal angle $\phi$ between the plane defined by exchanged photon and dijet system and the plane defined by the incoming and scattered lepton momenta in the $\gamma^*$-Pomeron rest frame. The shape of the $\phi$ distribution is sensitive to the production mechanism of diffractive exclusive dijets. The data favour fully perturbative calculations of the two-gluon exchange model over boson-gluon fusion mechanism, in which a partonic structure of the Pomeron is assumed.

## Tests of QCD factorisation

In the framework of QCD collinear factorisation the cross section for the diffractive DIS process factorises into universal diffractive parton distribution functions (DPDFs) and hard subrocess cross sections calculable within the perturbative approach. The universality of these DPDFs was sucessfully tested for hard diffractive DIS production. However, the factorisation breaking was observed for hard processes in diffractive hadron-hadron scattering where the predictions based on HERA DPDFs overestimate the data, at Tevatron by about one order of magnitude. The factorisation breaking is usually explained in terms of multiple scattering effects, which occur in the presence of beam remnants and lead to the destruction of the rapidity gap associated with the diffractive process. Rapidity gap survival effects can be investigated at HERA in hard diffractive dijet photoproduction.

Diffractive dijet cross sections in photoproduction and in DIS with a leading proton detected in the very forward proton spectrometer (VFPS) have been measured by H1 [2]. DIS measurements are complemented by measurements of dijet production with an associated large rapidity gap (LRG) [3]. The cross sections for the dijet production in diffractive DIS with VFPS and with LRG are found to be well described by the NLO QCD predictions using the H12006 Fit-B DPDF set. These results confirm the validity of QCD factorisation in hard diffractive DIS.

Using the VFPS data the double ratio of measured to predicted cross sections in photoproduction to the corresponding ratio in DIS has been determined and its value is equal to $0.55 \pm 0.10(\text{data}) + 0.02(\text{theor.})$, with no significant kinematic dependence. This result is consistent with previous H1 measurements based on the LRG method and indicates that QCD factorisation may be broken in diffractive dijet photoproduction. However, ZEUS does not observe factorisation breaking in diffractive photoproduced jets.

# High-energy resummation effects in Mueller-Navelet jets production at the LHC

Bertrand Ducloué[1], Lech Szymanowski[2], Samuel Wallon[3,4]

[1]Department of Physics, University of Jyväskylä, P.O. Box 35, 40014 University of Jyväskylä, Finland
[2]National Centre for Nuclear Research (NCBJ), Warsaw, Poland
[3]LPT, Université Paris-Sud, CNRS, 91405, Orsay, France
[4]UPMC Univ. Paris 06, Faculté de Physique, 4 place Jussieu, 75252 Paris Cedex 05, France

The high energy limit of QCD, described by the Balitsky-Fadin-Kuraev-Lipatov (BFKL) approach [1, 2, 3, 4], has been the subject of many studies since four decades. One of the most promising processes to study these dynamics at hadron colliders was proposed by Mueller and Navelet [5]. Consider two jets separated by a large rapidity interval, i.e. each of them almost fly in the direction of the hadron "close" to it, and with similar transverse momenta. Besides the cross-section, much information on the QCD high-energy resummation effects can be obtained when studying the azimuthal correlations of the two jets [6, 7]. In a pure leading order (LO) collinear treatment, the two jets should be emitted back to back since there is no phase space for (untagged) emission between them. This simple picture is of course corrected by the inclusion of radiative corrections. On the other hand, in the high-energy limit, the multiple emission of semi-hard gluons between these two jets is expected to modify dramatically this picture, leading to enhanced cross-sections and strong decorrelation effects. However, it is known that passing from a leading logarithmic (LL) to a next-to-leading-logarithmic (NLL) treatment in the BFKL framework can modify significantly this picture.

Technically, such a BFKL treatment involves two building blocks: the jet vertex, which describes the transition from an incoming parton to a jet, and the Green's function, which describes the pomeron exchange between the vertices. A complete NLL BFKL analysis of Mueller-Navelet jets (for more details, see refs. [8, 9]), including the NLL corrections both to the Green's function [10, 11] and to the jet vertex [12, 2, 14, 15, 16, 17, 18], showed that the NLL corrections to the jet vertex have a very large effect, of the same order of magnitude as the NLL corrections to the Green's function [19, 20], leading to a lower cross-section and a much larger azimuthal correlation [21]. However, these results are very dependent on the choice of the scales, especially the renormalization scale $\mu_R$ and the factorization scale $\mu_F$, in particular in the case of realistic kinematical cuts for LHC experiments [8]. This dependency can be reduced by using the Brodsky-Lepage-Mackenzie (BLM) prescription [22], adapted to the resummed perturbation theory à la BFKL [23, 24], to fix the renormalization scale. Such a full NLL BFKL analysis supplemented by the BLM scale fixing procedure has been performed [9], leading to a very satisfactory description of the most recent LHC data extracted by the CMS collaboration for the azimuthal correlations of Mueller-Navelet jets at a center-of-mass energy $\sqrt{s} = 7$ TeV [25]. Recently, it was shown that energy-momentum non conservation should not affect this NLL BFKL analysis significantly [26].

In the very large center-of-mass energy of the LHC, the contributions of partons with very small $x$ fraction of the protons are dominating, and their distributions are therefore enhanced. This is a typical situation where multipartonic scattering contributions could be sizable, in particular when dealing with small transverse momenta of the tagged jets (which hopefully will become accessible at CMS and ATLAS) for which the usual twist suppression is not valid [27]. In such a situation, the cross-section gets an additional contribution where each jet is emitted by a separate BFKL chain. The exchange of two BFKL ladders is presumably the dominant contribution for asymptotical values of $s$, and it would be of great interest to determine at which point it starts to be of the same order of magnitude as the exchange of a single ladder. Thus, the study of this process with transverse momenta of the tagged jets as low as possible would be a great opportunity towards investigating the effect of multipartonic contributions at small $x$.

# New results on forward jets within High Energy Factorization

Piotr Kotko[†]

*Penn State University, University Park, PA 16802, USA*

## High Energy Factorization and Off-shell Amplitudes

In the High Energy Factorization (HEF) approach for forward particle production, a cross section is expressed as a convolution of unintegrated gluon density (UGD), collinear PDFs, and off-shell gauge invariant tree-level partonic cross section. Generally, UGD undergoes BFKL-like evolution, possibly taking into account subleading corrections. Construction of off-shell gauge invariant matrix elements is achieved by taking the Lipatov's effective action, or one of the approaches [3, 2] which are more suitable for a numerical calculation. In the present contribution we discussed one more approach based on the Wilson lines [4], which is more general and aims at constructing gauge invariant Green's functions with arbitrary number of off-shell legs.

## Forward jets at LHC

Here we have concentrated on two recent studies within HEF: central-forward dijet production [4] and $Z_0$+jet production at LHC. The sample results for the azimuthal decorrelations are shown in Fig. 1. The UGD that has been used here is the one described in [5] and abbreviated here as KS. It utilizes BFKL type evolution with kinematic constraint, DGLAP corrections and nonlinear BK term. This UGD has been supplemented with a hard scale dependence via the Sudakov form factor as described in [4] (a similar model [6] leads to similar results for the observables under study). This Sudakov-type resummation turns out to be important to have a reasonable agreement with the data. For the $Z_0$+jet production we recover correctly the normalized distributions, while the absolute cross sections are significantly smaller than the experimental ones, which may give a hint that MPIs should be incorporated in HEF.

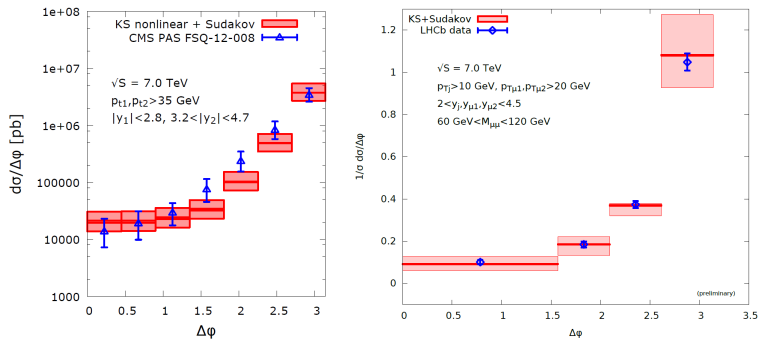

Figure 13: Azimuthal decorrelations for forward-central dijets (left) and Z0+jet production (right). The theory uncertainty bands come from the scale variation.

---

[†] in collaboration with A. van Hameren, K. Kutak and S. Sapeta

# The heavy quark impact factor and single heavy quark production

[a]Michal Deák, [b]Grigorios Chachamis, [a]Germán Rodrigo

[a]Instituto de Física Corpuscular, CSIC, Universitat de Valencia, E-46071 Valencia, Spain
[b]Instituto de Física Teórica UAM, CSIC and Universidad Autónoma de Madrid, E-28049 Madrid, Spain

## Motivation

The developments within the Balitsky-Fadin-Kuraev-Lipatov (BFKL) framework for the resummation of high center-of-mass energy logarithms [1] made possible the phenomenological analysis of high energy scattering processes within the $k_T$-factorization scheme. A key ingredient for studying high energy scattering processes within the $k_T$-factorisation scheme is the impact factor. The impact factors for gluons and massless quarks have been calculated at NL$x$ accuracy. The bottom quark became a suitable probe for QCD physics whereas the the LHC experiments are very efficient in tagging the $b$-quark. In addition the LHC could be considered to be a $b$-factory. Advantages connected with studying heavy quarks can be used to learn more about the small-$x$ physics. The NL$x$ impact factor for a massive quark in the initial state has been calculated in Ref. [4]. We may apply mass corrections calculated in [4] to calculate the cross section of single heavy quark production in proton-proton scattering at the LHC collision energies.

## Single heavy quark production

A good starting point for the solution of the above problem is to first calculate the cross section without the mass corrections. We can focus on a region of phase space where the mass corrections are relatively negligible, for $\mathbf{k}^2 \gg m_Q^2$. We apply the scheme used in [5]. We replace the photon impact factor in [5] with the quark impact factor to obtain the heavy quark-proton scattering cross section $\sigma_{pQ}$. To obtain the $pp$ scattering cross section we have to convolute $\sigma_{pQ}$ with the bottom quark parton distribution function. Using the framework above, we have calculated the transversal momentum distribution of the final state $b$-quark and compared it with a similar calculation done with the Monte Carlo program PYTHIA [6]. Both obtained cross sections agree well in the range of transversal momentum $30\ GeV < \mathbf{k} < 140\ GeV$ and rapidity $2 < y < 2.4$.

# Compact expressions for off-shell amplitudes

Andreas van Hameren

Institute of Nuclear Physics, Polish Academy of Sciences

Within *collinear factorization*, the momenta of the particles and partons associated with the hard scattering amplitude are on-shell, and the amplitude is defined applying the Lehmann Symanzik Zimmermann reduction prescription on a connected Green function. For a gauge theory like quantum chromo dynamics (QCD), this approach ensures the necessary gauge invariance. For *high-energy factorization* or $k_T$-*factorization* [1], however, the initial-state momenta need to be off-shell, and the above approach cannot be applied directly.

There exist a few approaches in literature to define and calculate scattering amplitudes with off-shell partons. One of them uses an effective action, which introduces new fields representing the off-shell partons with given longitudinal momentum associated with them [2]. They enter the QCD Lagrangian via terms similar to Wilson lines, ensuring gauge invariance of the eventual scattering amplitudes involving these fields.

In another approach [3], which is more suitable for automated numerical implementations, the initial-state off-shell partons are replaced by auxiliary on-shell quarks, and auxiliary on-shell quarks or photons are added to the final state. The on-shellness insures a gauge invariant definition of the amplitude, and the eikonal Feynman rules the auxiliary quarks turn out to follow lead to the same amplitudes as the the effective action approach.

In [4], finally, a manifestly gauge invariant and constructive definition of scattering amplitudes with an arbitrary number of off-shell external gluons was presented, by considering matrix elements of Fourier transforms of straight infinite Wilson line operators associated with the off-shell external gluons. In this work, it becomes particularly clear that the key to gauge invariance is the association of a *direction* $p^\mu$ with each off-shell external momentum $k^\mu$, which satisfies $p\cdot k = 0$. With each direction, an eikonal line is associated.

Using the latter approach as a staring point, it has been demonstrated in [5] how tree-level multi-gluon amplitudes with an arbitrary number of them off-shell can be calculated via Britto Cachazo Feng Witten (BCFW) recursion [6]. The BCFW method leads to very compact expressions with relatively little effort for helicity amplitudes with only on-shell partons. In [5] a straightforward generalization of BCFW recursion has been established, allowing again for compact expressions with relatively little effort. In particular, helicity amplitudes for up to 4 gluons have been calculated with up to 3 of them off-shell, and the MHV formulas have been shown to hold for arbitrary number of gluons with up to 2 of them off-shell. Further generalization to quarks and the calculation of all 5-parton helicity amplitudes with up to 2 of them off-shell is in progress.

# Report from the LHC Forward Physics Working Group

Timothy A. D. Martin

University of Warwick

## Introduction

The LHC Forward Physics Working group brings together expertise from the ALICE, ATLAS, CMS, LHCb, LHCf and TOTEM collaboration along with theoretical colleagues involved in research at forward rapidity. A CERN yellow report covering forward physics topics will be released in early 2015 and outline a near term forward physics programme at the LHC.

## Forward Physics

Forward physics topics span a range of cross sections from the many milibarn high energy $pp$ diffractive cross section - of importance to cosmic ray air-shower physics among others, to high mass and potentially exotic centrally produced states at small cross sections. The forward physics report covers both hard and soft diffractive, central exclusive and photon induced processes, cosmic ray and heavy ion physics along with processes sensitive to BFKL and saturation effects such as Mueller Navelet jets.

## Forward Detectors

The established forward proton tagging TOTEM experiment at CMS and ALFA group of ATLAS will operate in 2015 and beyond, continuing their primary remit of the measurement on the total cross section along with novel measurements of forward particle spectra.

Two new forward detector proposals, ATLAS Forward Physics and the CMS-TOTEM Precision Proton Spectrometer[1] are approved by their respective collaborations. These are designed to have a sensitivity of approximately $3\% < \xi < 12\%$ ($\xi$ is the fractional momentum loss of a driffractivly scattered proton) during standard LHC physics runs with squeezed beams ($\beta^* \sim 0.55$ m). This results in sensitivity to centrally produced systems with mass $M_X > 200$ GeV. The detectors will ultimately perform track reconstruction with 3D pixel detectors and integrate fast timing detectors based on time-to-digital chips or Čerenkov photo-detectors. 10 ps timing resolution will allow for $\sigma_z = 2$ mm central vertex determination in a pile up environment.

Forward scintillator shower counters are being installed upstream and downstream of the ALICE, CMS and LHCb detectors. These will allow for the detection and hence study or veto of low mass proton dissociation in analyses of diffraction, rapidity gaps and central exclusive production.

## Outlook

Improved proton tagging capabilities at the LHC will enable new studies of electroweak and QCD physics at high $Q^2$ with sensitivity to potential beyond Standard Model effects.

# Kinematic corrections to Quarkonium production in NRQCD approach

Sergey Baranov

P.N.Lebedev Institute of Physics
Russian Academy of Sciences

## Introduction and motivation

Quarkonium production in high energy hadrinic collisions is usually expressed in terms of non-relativistic QCD (NRQCD) factorization approach [1]. In its mathematical aspects, this approach is similar to the classical theory of multipole radiation; in practice, numerical calculations are always accompanied by an exordium that a $Q\bar{Q}$ pair produced in hard subprocess as color-octet emits (a number of) soft gluons changing the quantum numbers of the $Q\bar{Q}$ system without changing its energy-momentum. This is in obvious contradiction with confinement which prohibits the emission of soft colored quanta. Our note represents an attempt to estimate the numerical effect of final state radiation on the experimental observables if to replace an 'infinitely soft' gluon with a 'real' quantum of nonzero (of order $\Lambda_{QCD}$) energy-momentum.

## Results and conclusions

The simulation is organized as follows. First, we generate the production of a color-octet $c\bar{c}$ state in a usual perturbative way, but with a mass $m_{[8]}$ slightly higher than the nominal $J/\psi$ mass in order to preserve some phase space for the subsequent decay. As an example we tried $m_{[8]} = 3.4$ and 3.7 GeV. After that we generate a two-body decay $c\bar{c}[8] \to J/\psi + X$. We do not specify the nature of the state $X$, since the only property of importance is its mass $m_X$. For illustration we tried $m_X$=0, 150, 300, and 500 MeV.

The $p_t$ of the resulting $J/\psi$ is smaller than that of the originall $c\bar{c}$ pair, as part of the initial momentum is carried away by the quantum $X$. The visible kinematic effect (i.e., the ratio of the cross sections before and after emission $d\sigma(c\bar{c}[8])/dp_t$ and $d\sigma(J/\psi)/dp_t$) turned out to be very similar for all intermediate octet states ($^1S_0^{[8]}$, $^3S_1^{[8]}$, $^3P_0^{[8]}$, $^3P_1^{[8]}$, $^3P_2^{[8]}$), despite the individual spectra on their own may be rather different. The effect is nearly insensitive to the mass of the emitted quantum $m_X$ and is only sensitive to the mass difference between $m_{[8]}$ and $m_\psi$. The kinematic correction is large and, therefore, cannot be neglected. On the other hand, it flattens at high $p_t$, and so, can be effectively included into redefined nonperturbative matrix elements.

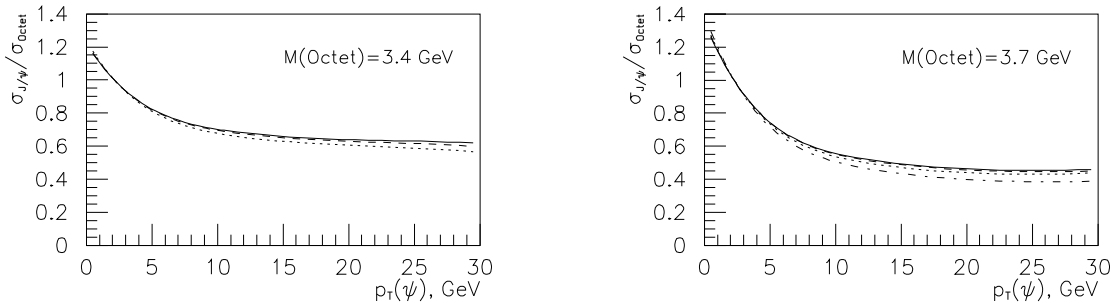

Kinematic correction factors for $p_t$ spectra due to final state radiation. The mass of the emitted system $m_X$ is set to 0 (solid curves), 150 MeV (dashed), 300 MeV (dotted), and 500 MeV (dash-dotted). The octet state mass is set to 3.4 GeV (left panel) and 3.7 GeV (right panel).

## References

[1] G. T. Bodwin, E. Braaten, and G. P. Lepage, Phys. Rev. D **51**, 1125 (1995); **55**, 5853(E) (1997).

# Unintegrated PDFs from low to high $x$: theoretical issues

I.O. Cherednikov

Universiteit Antwerpen, Belgium

Partonic content of nucleons beyond the collinear approximation is probed in *semi-inclusive reactions* with polarized and unpolarized hadrons, where the intrinsic transverse momenta and spin-orbit correlations of the partons are directly accessible [1]. To this end, collinear PDFs are replaced by the *transverse-momentum dependent* parton densities (TMDs) [2].

Fully inclusive processes are studied within the collinear QCD factorization framework making use of ($\mathbf{k}_\perp$-integrated) PDFs. DGLAP resummation is based on the strong ordering of emitted patrons in their transverse momenta, which suggests that the hard matrix elements are on-shell (e.g., the virtualities are much smaller than the hard scale) [3]. The DGLAP evolution sums up $\ln \mu^2/\Lambda^2$ logarithms.

Theoretical understanding of the semi-inclusive processes calls for updated factorization methods, which include various *unintegrated* PDFs.

*BFKL* method is directly applicable in the high energy or small-$x$ regime. In contrast to DGLAP, it suggests the strong ordering in the rapidities of emitted partons. The hard matrix elements are, therefore, allowed to be off-shell [4]. This method enables resummation of the $\ln 1/x$ logarithms and allows for nonlinear extensions.

*CCFM* approach makes use of the angular ordering of the emitted patrons, the hard matrix elements are also off-shell and the dependence on extra scale (maximal angle) is introduced. As such, this approach is not restricted to the small-$x$ regime but is valid as also for the large-$x$. It $\ln 1/x$ as well as $\ln 1/(1-x)$ logarithms [4].

*TMD* framework is constructed as a direct generalisation of the collinear picture. The operator definitions of TMD PDFs are formally valid in the whole range of $x$. However, beyond the tree-approximation a number of specific singularities emerges (the light-cone or rapidity divergences, mixed UV/rapidity terms, etc.), which calls for extra rapidity scale and yields more complicated, as compared to the collinear case, evolution equations [2, 5, 6].

Soft-Collinear Effective Theory *(SCET)* utilises the idea of the energy scale separation that results in the construction of an effective Lagrangian, the latter being used to obtain the TMD correlators. This approach is shown to be equivalent to the TMD scheme [6].

Given that the unintegrated PDF framework is aimed to provide a unique theoretical scheme for small- and large-$x$ experiments (EIC, JLab, RHIC, AFTER@LHC), the detailed comparison of the TMD evolution with DGLAP, BFKL, nonlinear small-$x$ and CCFM appears to be one of the most urgent tasks [7].

# Kinematical constraint and the CCFM equation

[a]Michal Deák, [b]Krzysztof Kutak

[a]Instituto de Física Corpuscular, CSIC and Universitat de Valencia, E-46071 Valencia, Spain
[b]Instytut Fizyki Jădrowej im. H. Niewodniczańskiego Radzikowskiego 152, 31-342 Krakow, Poland

## Introduction

In the high energy regime of hadronic scattering (the centre of mass energy is larger than any other present scale), perturbative approach to processes with high momentum transfer allows factorisation of the cross section into a hard matrix element and a gluon density [1]. In analogy with the BFKL approach the gluon density is a function of the longitudinal momentum fraction $x$ and transverse momentum $\mathbf{k}$ of a gluon. In addition dependence on a scale $p$ related to a hard process is introduced. For small densities the gluon density obtained in the above approach the gluon density obeys the CCFM equation [2]. The CCFM equation incorporates gluon radiation coherence by summing gluon ladder diagrams with rungs ordered in angle.

The angular ordering – an essential ingredient of the CCFM equation – introduces a constraint on kinematic variables in the integral on the right-hand-side of the equation. It turns out it is necessary to include also the kinematical constraint [3] into the CCFM and the BFKL equations as well since it was assumed in the derivation of both of them ensuring the gluon 4-momentum invariant to be dominated by $-\mathbf{k}^2$. The kinematic constraint causes suppression of the amplitude growth especially for low $|\mathbf{k}|$. An effect of suppression of the growth of the amplitude is achieved also by introducing saturation effects into the equation. The latter is achieved by including a negative quadratic term on the right hand side of the equation [4].

## Outlook

For the purpose of seeing the size of the corrections induced by the kinematical constraint and the non-linear term in the CCFM and the BFKL equation we solve it on a grid by iteration with the initial condition $\mathcal{E}_0(\mathbf{k}) = 1/(2|\mathbf{k}|)$.

The correction induced by the kinematical constraint is relatively big especially for small $\mathbf{k}$ for the BFKL and the CCFM equation as well and cannot be neglected.

The non-linear term causes the amplitude to be suppressed for $\mathbf{k}$ smaller than certain scale $Q_s$ which could be identified with the saturation scale. The size of the suppression depends on the constant parameter $R$ – coefficient in front of the quadratic term.

# Forward Drell-Yan cross-section beyond single scattering approximation

Tomasz Stebel, Leszek Motyka and Mariusz Sadzikowski

M. Smoluchowski Institute of Physics, Jagiellonian University,
S. Lojasiewicza str. 11, 30-348 Krakow, Poland

## Open problems

Forward Drell-Yan (DY) process at the LHC is a sensitive tool for investigating QCD effects at small Bjorken-$x$ ($x \simeq 10^{-6}$ and moderate energy scales). The LHCb experiment is expected to measure DY lepton-antilepton mass $M$ of about 2.5 GeV. In this kinematical domain higher twists of Operator Product Expansion (OPE) may introduce sizeable corrections to the leading twist approximation.

In the standard description of the DY process an angular distribution of lepton pairs is described by four structure functions $W_L$, $W_T$, $W_{TT}$, $W_{LT}$. Higher twist analysis of forward the DY is based on the OPE of these functions. The power suppression in hard scale of higher twists is compensated by small-$x$ growing of matrix elements. In addition, combination $W_L - 2W_{TT}$ vanishes at leading twist-2 up to next-to-next-to-leading order of QCD corrections (this is known as the Lam-Tung relation). This makes from $W_L - 2W_{TT}$ a very promising observable for future experimental searches of higher twists.

In the high energy limit in QCD, relevant for forward Drell-Yan process at the LHC, the scattering amplitudes may be efficiently computed using the $k_T$-factorization framework and the color dipole picture. In our approach we expressed structure functions $W_i$ as a Mellin convolutions $W_i \sim \int_C ds\, \tilde{\sigma}(-s)\hat{\Phi}_i(s)$ where $\tilde{\sigma}$ and $\hat{\Phi}_i$ are Mellin transforms of the dipole cross-section and an impact factor, respectively. Impact factors are calculated from perturbative QCD at the leading order and hence are universal (do not depend on the model) and their form is one of the main results of our calculations. On the other hand the cross-section $\tilde{\sigma}$ needs to be modeled.

As a first step towards understanding the higher twist structure of the forward DY scattering we consider the simplest case of the eikonal dipole cross-section used in the Golec-Biernat–Wüsthoff (GBW) approach. Using this cross-section we perform an explicit analytic twist decomposition of the forward DY structure functions. The GBW model provides an efficient description of the small-$x$ data. One should note however that up to now the data probe the proton structure efficiently only at the leading twist. Hence, current theoretical predictions of higher twist effects are uncertain.

## Outlook

Since our results for the twist expansion of the forward DY structure functions are model dependent one needs also to provide dedicated predictions for the LHC within other reasonable schemes. Balitsky-Fadin-Kuraev-Lipatov (BFKL) and Balitsky-Kovchegov (BK) approximations obtained using small-$x$ resummations are the most promising ones. The next step in our program of theoretical analysis of the LHC higher twist will be a data oriented study containing numerical predictions for the higher twist corrections to forward DY within different models (such as GBW, BFKL, BK).

For results, more detailed description and references see [1].

Chapter:
# MPI in collisions with nuclei

# Outlook on MPI and the Heavy Ions

P. Bartalini

This brand new working group deals with the interplays between the MPI and the Heavy Ions research lines.

The LHC RUN I highlights in this field include the unexpected observation of long-range, near-side angular correlations in small systems (i.e. pp and pPb collisions) [1, 2] as well as the systematic studies of high multiplicity events, the latter pointing to a fundamental role of rather soft MPI in driving the final states with large multiplicities: the ALICE event shape analysis [Eur.Phys.J. C72 (2012) 2124] demonstrates that these events look more spherical than predicted by the general purpose QCD models, disfavoring the presence of well-pronounced jet structures. This is basically confirmed by the CMS analysis focusing on the study of jets and Underlying Event (UE) activity in large multiplicity events [3]. ALICE also launched a systematic investigation of event yields versus charged multiplicity, Nch, [4] that - at least in pp - corroborates the trivial picture in which Nch and all the event yields are proportional to the number of MPI. In pPb, the observed discrepancies to this rule, point to the need of improving the centrality determination in these asymmetric interactions [5]; this requires a strict collaboration between experimentalist and theorist [M. Strikman, this proceedings].

Progress in Run 2 is expected by the adoption of large multiplicity triggers at low luminosity: as far as pPb interactions are concerned all the experiments are expected to improve while only ALICE will take large low luminosity pp samples at 13 TeV, ending up with at least one million of events with average multiplicity of one order of magnitude higher than the minimum bias average. Progress should also arise from the adoption of new experimental techniques that allow separating jetty from not jetty events using the event shapes [6]. Possible variations of the hadrochemistry in large multiplicity events should be investigated in detail; alternative explanations to this flow-like phenomenon may be provided by MPI models with color reconnections [G. Paic, 5th MPI@LHC]. A significant progress in understanding flow-like effects in small systems with large multiplicities in the final states is expected by the adoption of models describing MPI in an homogeneous way in both pp and pPb interactions [C. Bierlich, this proceedings].

A lot of emphasis has been put in the possible observation of the double parton scattering (DPS) in pPb events [D. Treleani, 5th MPI@LHC and most of the theoretical contributions in this section of the proceedings]. Indeed Ð due to anti-shadowing effects - the DPS signal/background ratios in pPb are expected to be enhanced with respect to the corresponding final states in pp. However one needs to be very careful in taking into account the LHC luminosity figures: in the LHC RUN I all the LHC experiments got around 10 nb-1 at centre-of-mass energies per nucleon pair of 5.02TeV; given the current constraints, the RUN II plans are to collect data at the same energy integrating a significantly higher statistics that in any case will not go beyond 100 nb-1. This means that the DPS measurements in pPb should select final states with large cross section; the significant pp progress [arXiv:1410.6664 and DPS section of these proceedings] suggests to focus on final states with multiple heavy flavors production.

# Evidence for $x$-dependent proton color fluctuations in pA collisions at the LHC

M. Alvioli[1], B. A. Cole[2], L. Frankfurt[3], M. Strikman[4]

[1] CNR-IRPI, via Madonna Alta 126, I-06128, Perugia, Italy
[2] Columbia University, New York, NY 10027, USA
[3] Tel Aviv University, Tel Aviv, Israel
[4] 104 Davey Lab, The Pennsylvania State University, University Park, PA 16803, USA

In pA collisions at collider energies a projectile stays in a frozen configuration over the distances which by far exceed the nuclear diameter. As a result the proton coherently interacts with nucleons along its impact parameter. In QCD a nucleon is build of quark-gluon configurations of different transverse size which are expected to interact with different strength, leading to fluctuations of the global strength of the projectile interactions. Also, configurations of smaller size are expected to have a reduced gluon field leading to a correlation of soft and hard interactions. To describe these effects it is convenient to introduce a probability distribution for the projectile to interact with strength $\sigma$, $P_N(\sigma)$. $P_N(\sigma)$ is strongly constrained by the diffractive $pp$ data, the behavior of $P_N$ for $\sigma \to 0$ expected in pQCD, etc [1]. We developed a Monte Carlo procedure to describe these effects in soft collisions and collisions with a hard trigger taking into account the difference of the transverse scales of hard and soft interactions [2, 3]. The prediction that the distribution over the number of wounded nucleons should be broader than the Glauber model expectation was confirmed at the LHC [4]. We argue that the violation of the Glauber approximation for the dependence of the rate of forward jet production on centrality observed in $pA$ collisions at the LHC provides the first experimental evidence that parton configurations in the projectile proton containing a parton with large $x_p$ interact with a nuclear target with a significantly smaller than average cross section, and have smaller than average size. Implementing effects of such fluctuations of the interaction strengths and using the ATLAS analysis [4] of the dependence of the hadron production at backward rapidities on the number of wounded nucleons, we make quantitative predictions for the centrality dependence of the jet production rate as a function of the interaction strength $\sigma(x_p)$. For $x_p = 0.6$ we find $\sigma(x_p) \sim \sigma_{tot}(pp)/2$ [5] which has implications on the origin of the EMC effect. Future studies of dijet production along these lines would allow to investigate the global 3D structure of the nucleon and to trigger on particular configurations which interact with a larger-than-average strenght.

# Double-scattering production of two $\rho^0$ mesons and four pions in heavy ions UPCs

Mariola Kłusek-Gawenda, Antoni Szczurek

Institute of Nuclear Physics Polish Academy of Sciences

## Open problems

We have calculated cross sections for double-$\rho^0$-meson as well as four-pion production in exclusive ultraperipheral ultrarelativistic heavy ion collisions via a double-scattering mechanism of single $\rho^0$ photoproduction. The calculations have been done in the impact parameter space. We have included the smearing of $\rho^0$ meson masses and used the latest ALICE data to parametrize the Breit-Wigner amplitude and the Drell-Söding type continuum term. These data allows to consider $\rho^0\rho^0$ or corresponding $\pi^+\pi^-\pi^+\pi^-$ double photoproduction in broader range of center of mass energy of the four pion system. Our results for single-$\rho^0$ production agree well with the STAR and ALICE experimental data.

The produced $\rho^0$ mesons decay into charged pions, with almost 100% probability, gives large contribution to exclusive production of the $\pi^+\pi^-\pi^+\pi^-$ final state. We have made a comparison of four-pion production via $\rho^0\rho^0$ production ($\gamma\gamma$ fusion [1] and nuclear photoproduction [2]) with experimental data measured by the STAR Collaboration for the AuAu→AuAu$\pi^+\pi^-\pi^+\pi^-$ reaction. The theoretical predictions have similar shape in four-pion invariant mass as the distribution measured by the STAR Collaboration, but exhaust only about 20% of the measured cross section. The missing contribution can come from excited state of $\rho^0(770)$ (e.g. $\rho^0(1700)$ resonance) and its decay into four charged pions. A separation of the double-scattering, $\gamma\gamma$, $\rho^0(1450)$ and $\rho^0(1700)$ production and decay mechanisms seems very important. In general, transverse momentum of each of the produced $\rho^0$'s in the double-scattering mechanism is smaller than in the other mechanisms. As a consequence, the pions from the decay of $\rho^0$'s from the double-scattering mechanism are produced dominantly back to back in azimuthal angle. This could be used to enhance the purity of the experimental sample as far as double-scattering mechanism is considered. At large rapidity separations between two $\rho^0$'s and/or large $\pi^+\pi^+\pi^+\pi^-$ rapidity separations, the double-scattering contribution should dominate over other contributions. The identification of the dominance region seems difficult, if not impossible, at RHIC. We plan a separate careful analysis devoted to the ALICE experimental conditions. It would be interesting if different mechanisms discussed in this talk could be separated and identified experimentally. This requires, however, rather complicated correlation studies for four charged pions.

## Outlook

We have concentrated on the process with final nuclei in the ground state. It is very difficult, if not impossible, to measure such very forward/backward nuclei. The multiple Coulomb excitations (Ref. [3]) associated with $\rho^0\rho^0$ production may cause additional excitation of one or even both nuclei to the giant resonance region and subsequent emission of neutrons. We plan a detailed study of these processes in the future.

# Transverse momentum dependent gluon density at low $x$

A.A. Grinyuk[1], A. V. Lipatov[1,2], G. I. Lykasov[1], N. P. Zotov[2]

[1] Joint Institute for Nuclear Research - Dubna 141980, Moscow region, Russia
[2] Skobeltsyn Institute of Nuclear Physics, Lomonosov Moscow State University, 119991, Moscow, Russia

We present results on the application of the transverse momentum $\mathbf{k}_T$ dependent gluon density (TMD) at low $x$, which based on our previous study [1]. We match this TMD at low transverse momenta $|\mathbf{k}_T|$ and starting scale $Q_0^2 = 1 - 2$ GeV$^2$ to the exact solution of the BFKL equation outside of the saturation region at large $|\mathbf{k}_T|$ obtained in [2]. Then, to extend this TMD at higher $Q^2$ we use the Catani-Ciafoloni-Fiorani-Marchesini (CCFM) evolution equation solution. The inclusion of the CCFM evolution results in a large increase of the TMD magnitude. The use of new TMD allows us to describe satisfactorily the LHC data on the heavy flavour jet production and especially the correlation between two $b$-jets produced in $pp$ collisions at low $x$ and large $|\mathbf{k}_T|$. In Fig.1 (left) the rapidity $y$ distribution of $b$-jets produced in the $pp$ collision at $\sqrt{s}$ =8 TeV is presented. The distribution over the difference of the azimuthal angles $\Delta\phi$ between similar two $b$-jets is presented in Fig.1(right).

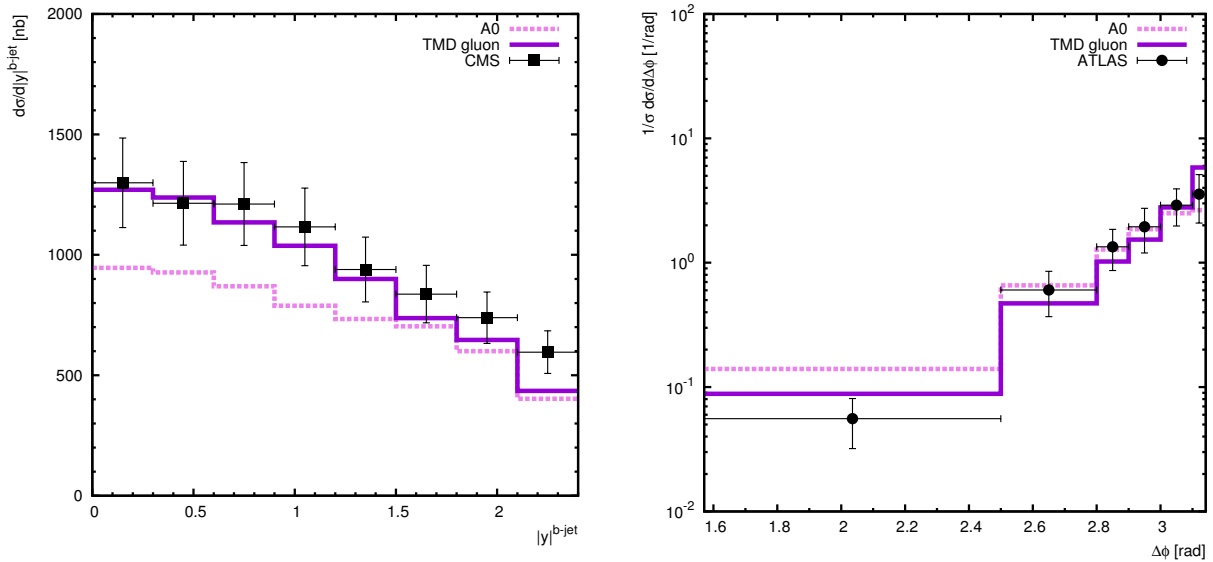

Figure 14: The $y$-distribution of $b$-jets produced in the $pp$ collision at $\sqrt{s}$ =8 TeV compared to the CMS data [3] (left). The distribution over the difference of the azimuthal angles $\Delta\phi$ between similar two $b$-jets compared to the ATLAS data [4] is presented on the right. The solid hystograms correspond to our new TMD, whereas the dashed ones correspond to the TMD suggested in [5] (set $A0$).

# Double parton interactions in $\gamma p, \gamma A$ collisions in the direct photon kinematics.

B. Blok[1], M. Strikman[2]

[1] Department of Physics, Technion – Israel Institute of Technology, Haifa, Israel
[2] Physics Department, Penn State University, University Park, PA, USA

[1] Joint Institute for Nuclear Research - Dubna 141980, Moscow region, Russia
[2] Skobeltsyn Institute of Nuclear Physics, Lomonosov Moscow State University, 119991, Moscow, Russia

*Multiple hard parton interactions* (MPI) started to play an important role in the description of the inelastic $pp$ collisions at the collider energies. Extensive theoretical studies were carried out in the last decade, both for pp collisions and for $pA$ collisions. In particular it was realized that MPI in pp collisions are expressed through convolution of new objects-two parton GPD [1], and determined by the two different mechanisms $1\otimes 2$ and $2\otimes 2$ (see refs. [1, 2, 3, 4] and [5] for details and notations.) Presence of two mechanisms and limited knowledge of the nucleon multiparton structure makes a unique interpretation of the data rather difficult. Hence we proposed [5] to study the MPI process of $\gamma p(A)$ interaction with production of four jets in the kinematics where two jets carry most of the light cone fraction of the photon four momentum - direct photon mechanism. In this process the $2 \otimes 2$ mechanism is absent and the only process which contributes is an analog of the $1 \otimes 2$ process. Since in the proposed kinematics the contribution of the resolved photon is strongly suppressed the cross section in the leading log approximation (LLA) i.e. summing leading collinear singularities is expressed through the integral over two particle GPDs, $_2D(x_1, x_2, Q_1^2, Q_2^2, \Delta)$, introduced in [1]. This is in difference from the case of $pp, pA$ scattering where $2\otimes 2$ contribution is proportional to a more complicated integral with the integrant proportional to the product of two double parton GPDs.

Our main result is that processes with direct photon in photon proton collisions considered in [5] provide a golden opportunity for the model independent determination of the double parton distributions $_2D$, free of the ambiguities inherent in $pp/pA$ scattering [3]. We consider the process of the interaction of the real/quasireal photon with proton with production of two pairs of hard jets in the back to back kinematics with each dijet consisting of a heavy (charm) quark and gluon jets (see Fig. 1). We focus on the production of charm to suppress the contribution of the resolved photons.

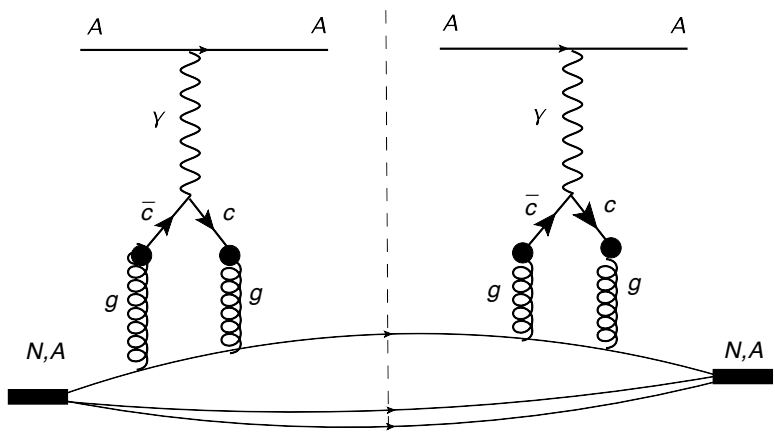

Figure 15: MPI two dijet photoproduction -$Ap$

In the discussed process a $c\bar{c}$ pair is produced in the photon fragmentation region, while two gluon jets are created predominantly in the target region, so that there is a large rapidity gap between the gluon and quark jets. The gluon and c-quark jets are approximately balanced pair vice. The cross section of the analogous process in $pp$ collisions is influenced by parton correlations in both nucleons participating in the process, while in the case of the photon the cross section depends only on the integral over one wave function. The reason is that the process involves only one GPD from the nucleon, while the upper part of the diagram 1 is determined by the hard physics of the photon splitting to $Q\bar{Q}$ pair in an unambiguous way. It does not involve the scale $Q_0^2$ that separates perturbative and nonperturbative correlations in a nucleon. Thus the cross section of such a process is directly expressed through the nucleon double GPD. Hence the measurement of the discussed cross section would allow to perform a nearly model independent analysis of DPI in $pp$ scattering.

Here we consider the MPI rates for all three types of processes mentioned above, $\gamma p, AA, pA$. We will restrict ourselves to the kinematics $x_1, x_2 > 0.2$, thus guaranteeing the dominance of the direct photon contribution (For this cutoff the direct photons contribute 60% of the dijet cross section). For a lower $x_i$ cutoffs the relative contribution of the direct photon mechanism rapidly decreases for transverse momenta under consideration.

We see that for the LHeC collider energies $\sqrt{s} = 1.3$ TeV the rate of the discussed reaction will be very high: $2 \cdot 10^9$ events per $10^7$ s (one year of running) for the luminosity $10^{34}$ cm$^{-2}$s$^{-1}$ and $p_t > 5$ GeV. The relative rate of MPI to dijets is found to be 0.045%. Moreover, as it is clear from Fig. 5 below, we shall have a large MPI rate up to $p_t$ of order 17 GeV.

A large number of events in the discussed kinematics was produced at HERA: $\sim 1.2 \cdot 10^5$ for the total luminosity 1 fb$^{-1}$. It would be interesting to reanalyze the HERA data in the direct photon kinematics with the purpose of identifying MPI events.

Another way to observe the discussed process in the near future maybe possible - study of MPI in the ultraperipheral $pA, AA$ processes at the full LHC energy. For example, for $p_t > 5$ GeV, we have $\sim 6 \cdot 10^4$ events for $AA$, and $\sim 6.6 \cdot 10^3$ events for $pA$ scattering where we used luminosities $\sim 10^{27}$ ($AA$), $10^{29}$ cm$^{-2}$s$^{-1}$ ($pA$) and running time of $10^6$ s. In the discussed kinematics MPI events constitute $\sim 0.04\%$ ($\sim 0.02\%, \sim 0.0125\%$) of the dijet events for $AA, pA$ collisions respectively for the same jet cutoff. These fractions decrease rather rapidly with $p_t$ increase.

# DPS cross sections in proton-nucleus and nucleus-nucleus collisions at the LHC

David d'Enterria[a], Alexander M. Snigirev[b]

[a] CERN, PH Department, 1211 Geneva, Switzerland
[b] Skobeltsyn Institute of Nuclear Physics,
Moscow State University, 119991 Moscow, Russia

The derivation of simple generic expressions to compute double-parton scattering (DPS) cross sections in high-energy proton-nucleus [1] and nucleus-nucleus [2] collisions, as a function of the corresponding single-parton hard cross sections and of the event centrality, is presented. The large transverse parton density in nuclei results in a high probability of having two truly hard scatterings in p-A and A-A collisions. The DPS cross sections in p-Pb are found to be enhanced by a factor of $3A \simeq 600$ compared to those in p-p collisions at the same energy. The DPS cross sections in Pb-Pb are found to be enhanced by a factor of $A^{3.3}/5 \simeq 9 \cdot 10^6$ to be compared with the $A^2 (\simeq 4 \cdot 10^4)$ enhancement of single-parton scatterings.

The formalism was applied to estimate the DPS cross sections for quarkonium and electroweak-boson pair production in Pb-Pb and p-Pb collisions at the LHC. Table 1 collects these results [3, 4]. Processes such as double-$J/\psi$, $J/\psi + \Upsilon$, $J/\psi$+W, $J/\psi$+Z, double-$\Upsilon$, $\Upsilon$+W, $\Upsilon$+Z, and same-sign WW production have large cross sections and visible event rates for the nominal LHC luminosities. The study of such processes in p-Pb can help determine the effective nonperturbative parameter (so-called effective cross section $\sigma_{\text{eff}}$) characterising the transverse parton distribution in the nucleon. Double–$J/\psi$ and double-$\Upsilon$ production in Pb-Pb provide interesting insights on the event-by-event dynamics of quarkonia in hot and dense strongly-interacting medium.

| System | | $J/\psi + J/\psi$ | $J/\psi + \Upsilon$ | $J/\psi$+W | $J/\psi$+Z | $\Upsilon + \Upsilon$ | $\Upsilon$+W | $\Upsilon$+Z | ss WW |
|---|---|---|---|---|---|---|---|---|---|
| Pb-Pb | $\sigma$ | 210 mb | 28 mb | 500 $\mu$b | 330 $\mu$b | 960 $\mu$b | 34 $\mu$b | 23 $\mu$b | 630 nb |
| 5.5 TeV | $N$ | ~250 | ~340 | ~65 | ~14 | ~95 | ~35 | ~8 | ~15 |
| p-Pb | $\sigma$ | 45 $\mu$b | 5.2 $\mu$b | 120 nb | 70 nb | 150 nb | 7 nb | 4 nb | 150 pb |
| 8.8 TeV | $N$ | ~65 | ~60 | ~15 | ~3 | ~15 | ~8 | ~1.5 | ~4 |

Table 2: DPS production cross sections of double-$J/\psi$, $J/\psi + \Upsilon$, $J/\psi$+W, $J/\psi$+Z, double-$\Upsilon$, $\Upsilon$+W, $\Upsilon$+Z, and same-sign WW in Pb-Pb and p-Pb at the LHC. The corresponding DPS yields, after (di)lepton decays and acceptance+efficiency losses, are given for 1 nb$^{-1}$ and 1 pb$^{-1}$ respectively.

# Transverse momentum spectra of light-flavor hadrons measured with ALICE in pp, p–Pb and Pb–Pb collisions at the LHC

Jacek Otwinowski

Institute of Nuclear Physics
Polish Academy of Sciences

## Open problems

Measurements of p–Pb collisions at the LHC has provided a lot of surprising results. In particular, so-called ridge structure was seen in two-particle correlations measured in high-multiplicity p–Pb events [1], which is similar to that in Pb–Pb collisions originating from collective expansion of dense medium (hydrodynamical flow). The ridge structure modulation in azimuth is described in terms of Fourier coefficients [2], indicating that the collective flow might develop in such system. Evidence for the flow has been further corroborated from the particle-identification measurements of the ridge [3], showing the characteristic mass ordering in $v_2$ vs $p_T$ for pions, kaons and protons. However, the measurements of the average $p_T$ as a function of the charged particle multiplicity [4] indicate that the collective phenomenon seen in p–Pb collisions might be due to the color reconnection. The same conclusion can be drown from the interpretation of the Blast-wave fit to the identified particle $p_T$ spectra [5]. Moreover, the nuclear modification factors measured for charged particles [6] showed no suppression at high-$p_T$, indicating that the properties of matter produced in p–Pb and Pb–Pb collisions are different. In addition, the p–Pb system size determined using Bose-Einstein correlations of three pions [7] is much smaller (35-55%) than size of the Pb–Pb system, measured at the same particle multiplicities.

## Outlook

These findings hint at potentially novel mechanisms in collisions of small systems which are far from being understood theoretically. Several authors describe the results in the context of hydrodynamics [8], but also saturation models successfully describe some of the measurements [9]. Comparison of measurements in pp, p–Pb and Pb–Pb collisions at the same energy $\sqrt{s_{NN}} = 5$ TeV, panned in 2015–2018, will help to reveal the properties of matter produced in small systems.

# Finite $N_c$ effects in $pp$ and $AA$ Monte Carlo event generation with DIPSY

Christian Bierlich,

Institute of Astronomy and Theoretical Physics, Lund University

The BFKL based event generator DIPSY [1], works by dynamically building up initial states of collisions (of hadrons or nuclei) in a partonic cascade in transverse space and rapidity. After interaction, the real, outgoing partons are evolved through final state radiation in ARIADNE 5 [2], and hadronization with the string model implemented in PYTHIA 8 [3], to produce full, exclusive final states. Energy conservation and a number of subleading effects are added, notably finite $N_c$ effects in FSR and hadronization:

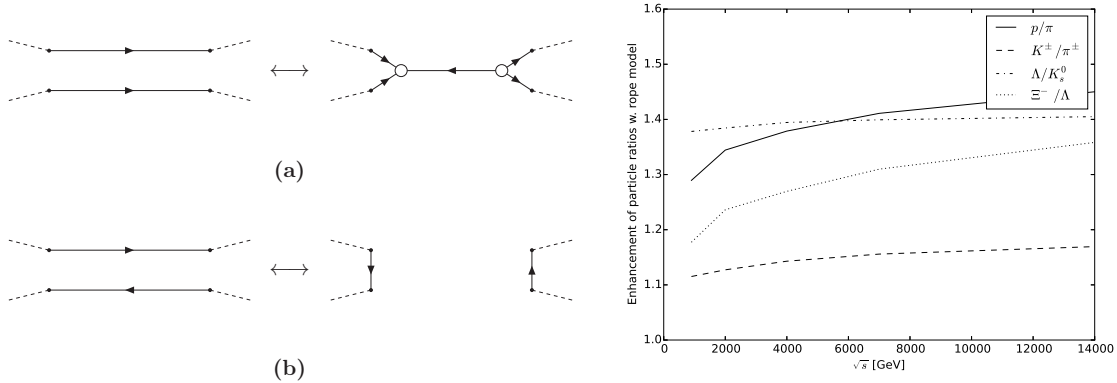

Figure 16: Left: reconnection of **(a)** two parallel dipoles, introducing two string junctions, and **(b)** two antiparallel dipoles, eliminating the string between them, using a swing mechanism. Right: $\sqrt{s}$-dependence (in $pp$) of integrated $p^\pm/\pi^\pm$, $K^\pm/\pi^\pm$, $\Lambda/K_s^0$ -and $\Xi^-/\Lambda$-ratios, as introduced by finite $N_c$ effects. The figure shows the enhancement in the rope model relative to ordinary string production, *i.e.* rope/string.

1. Allowing color compatible dipoles, with opposite color flow (see fig. 16 (left, **(b)**)), to reconnect with each other, with a probability:

$$\frac{d\mathcal{P}_s}{d\rho} \propto \frac{(\vec{p}_1 + \vec{p}_2)^2 (\vec{p}_3 + \vec{p}_4)^2}{(\vec{p}_1 + \vec{p}_4)^2 (\vec{p}_3 + \vec{p}_2)^2}.$$

2. Allowing overlapping strings to recombine in a random walk process, hadronizing the resulting "ropes" with a higher string tension, or make baryons of the resulting *junction structures* from dipoles with parallel color flow, see fig. 16 (left, **(a)**). Resultingly the strange/light hadron and baryon/meson ratios gets an $\sqrt{s}$-dependence (see fig. 16 (right)), which corresponds well to observations.

The finite $N_c$ effects introduced to DIPSY holds promise for explaining flavour ratios in $pp$ as well as $pA$ and $AA$, as the formalism of the model is identical in all three cases.

Work supported in part by the MCnetITN FP7 Marie Curie Initial Training Network, contract PITN-GA-2012-315877, and the Swedish Research Council (contracts 621-2012-2283 and 621-2013-4287).
In collaboration with Gösta Gustafson and Leif Lönnblad.

# Constraints for small-$x$ gluon nPDFs from p+Pb collisions at the LHC

Ilkka Helenius

Department of Astronomy and Theoretical Physics, Lund University

## Open problems

The current nuclear parton distribution function (nPDF) fits are mainly constrained by the deep inelastic scattering and Drell-Yan data (see e.g. ref. [1] for a recent review). These data provide strong constraints for quark distributions at $x > 0.01$ but gluon distributions are constrained only indirectly. Some analyses utilize also data for inclusive pion production in d+Au collisions at RHIC which provides further constraints for gluon distributions but as the kinematic reach is rather limited, the gluons remain unconstrained at $x < 0.01$.

In ref. [2] we studied how the gluon nPDFs at small values of $x$ could be constrained using the p+Pb collisions at the LHC. Special emphasis were on quantifying the different regions of $x$ that are probed with inclusive hadrons and direct photons at different rapidities according to NLO pQCD. In the case of hadron production we found that due to the convolution of the partonic spectra with the fragmentation functions the relation between the partonic kinematics and the final state hadrons is significantly smeared. In addition to fragmentation mechanism the direct photons can be produced also directly at the hard scattering. The presence of this so-called prompt component makes the direct photons more sensitive to the small-$x$ physics than inclusive hadrons at the same values of transverse momentum $p_T$ and pseudorapidity $\eta$.

In addition to the nPDF studies these results can be applied also to seek for deviations from the collinear factorization based calculations. Large deviations from the pQCD predictions at certain $p_T$ and $\eta$ would hint for breaking of the factorization and the values of $x$ in which this takes place could be estimated from the presented $x$ distributions. An example of such a factorization breaking effect is the gluon saturation built into the Color-Glass-Condensate -framework, see e.g. ref. [3] for related applications

## Outlook

In order to reduce the uncertainties in the current nPDFs, measurements with better than 10% accuracy are required. The required accuracy could be achieved by installing a forward electromagnetic calorimeter (FoCal) to the ALICE detector which have been considered as a possible upgrade of ALICE experiment [4]. Also LHCb experiment has potential to measure electroweak cross sections in nuclear collisions at large rapidities as demonstrated e.g. in ref. [5]. The proposed measurements at forward rapidities would serve also as a further test of factorization in nuclear collisions in a so far unexplored kinematic region.

Work supported by the MCnetITN FP7 Marie Curie Initial Training Network, contract PITN-GA-2012-315877.

# Summary and Conclusions

The Sixth International Workshop on Multiple Partonic Interactions at the Large Hadron Collider (MPI@LHC2014) is the latest installment of a series of very successful workshops dedicated to the understanding of the dynamics behind Multiple Partonic Interactions (MPI). As in previous editions of the MPI@LHC series, the quest for an improved description of the dynamics behind MPI has focused the aim of this workshop.

Once again, the workshop provided an unique setting that promoted active interaction between the experimental and theoretical communities. It encouraged the discussion of many of the different aspects of MPI, identified common concepts between apparently different lines of research and evaluated the impact of MPI on the LHC physics program.

The experimental collaborations have reported on measurements that show evidence of MPI effects in pp, pA and AA collisions. These effects can be noticed in several scenarios defined experimentally, such as minimum bias and underlying event measurements in pp collisions, as well as in many observables inspired by the phenomenology of MPI for pp, pA and AA collisions. At the heart of the experimental discussion were the recent advances on direct measurements of double-parton scattering (DPS) and on strategies to extract $\sigma_{eff}$ with the aid of phenomenology models implemented in Monte Carlo (MC) generators. It is also noticeable that there has been a growing interest on studies which look for MPI effects in events with topologies characterized by forward particle production and/or rapidity gaps.

The workshop has also debated some of the modern theoretical considerations and the modeling of MPI in MC event generators. Theoretical studies on unintegrated PDFs, double parton distribution functions, double parton cross-section predictions, electroweak boson production in DPS, correlations in particle pair productions and final state topologies with forward tagged particles, jets and/or rapidity gaps are just some examples of the work that has been presented at the workshop. It is also worth mentioning that since the first meeting of this workshop series, there have been many advances on MC generator development and tuning which originated from discussions held at the MPI@LHC workshops - this edition included - and are now part of a recognizable legacy of these meetings.

Looking ahead to 2015 we can only be very excited with the possibilities the restart of LHC will bring to the MPI community. The Run II of the LHC is expected to deliver data on proton-proton collisions at a centre-of-mass energy of 13 TeV as well as new collisions at high energies with heavy ions. In this context of being confronted with the dawning of a new era while still harvesting many lessons from the LHC Run I cycle, the workshop participants have debated and agreed on a few points we should document here for future reference when we reconvene in 2015. These include:

- Experimental collaborations are strongly encouraged to define a set of "common measurements" with event selection and kinematic cuts which can be performed by as many experiments as possible. Minimum bias and underlying event measurements are key targets for this exercise. The set of common measurements produced with Run I data and documented in the "Minimum bias and Underlying Event Working Group of the LPCC"[1] has been identified as a starting point which will be proposed to the experiments.

- It is also recommended that a common MC tuning should be used as a reference by the experimental collaborations, the MC authors and users who will be developing new models or tunings. The PYTHIA8 - Monash tune was chosen as the MC tune to be used as the reference.[2]

---

[1] The LHC Physics Center at CERN (LPCC) - Minimum Bias & Underlying Event Working Group. Documentation and plots for the common measurements produced with the LHC's Run I data is available at this web-address: *http://lpcc.web.cern.ch/LPCC/index.php?page=mb_ue_wg_docs*

[2] P. Skands, S. Carrazza, J. Rojo, Eur. Phys. J. C 74 (8) (2014) 3024.

- Isolating the DPS signal at the LHC remais quite challenging and usually implies on the reliance of model dependent corrections in order to interpret the measurements. There is a need to re-think the observables used to perform a direct measurement of DPS at the LHC. It is also recommended that, where possible, experimental collaborations seek to use consistent (if not common) definitions of phase-space selection cuts and background templates.

- The range of experimental scenarios traditionally investigated in the search for MPI has been broadened. MPI is relevant for measurements done with data collected from pp, pA and AA collisions. Recent theory studies indicate that stronger experimental signatures of DPS in pp collisions are predicted to be observed in topologies which include final states with forward tagged particles, forward jets and/or rapidity gaps. It is also recommended that for pp, pA and AA collisions, new analyses on high-multiplicity final states, jet activity rates, correlations in particle pair production (double open charm, J/$\Psi$ pair and electroweak boson pair, to mention a few) should be further pursued with the LHC's Run II data.

Many other advances and open problems were reported at the workshop and the reader will certainly have found these described in the various contributions collected in these proceedings.

In 2015, the $7^{th}$ International Workshop on Multiple Partonic Interactions at the LHC - MPI@LHC2015 - will be hosted by the Intenational Centre for Theoretical Physics, in Trieste, Italy. On behalf of the International Advisory Board for the MPI@LHC Workshop Series I would like to invite all those interested in advances on MPI to come and join us at the next edition of this workshop. We look forward to meeting all our MPI friends at Trieste. The year 2015 promises to bring us many new results and very interesting physics indeed!

*Arthur M. Moraes*
*Chairman of the International Advisory Board for the MPI@LHC Workshop Series*